\documentclass{aa}
\usepackage[dvips]{graphicx}
\usepackage{txfonts}
\usepackage{longtable,lscape}
\usepackage{natbib}
\usepackage{cases}
\usepackage{subfig}
\usepackage[]{units}
\bibpunct{(}{)}{;}{a}{}{,}
\usepackage{color}
\usepackage{url}
\def\be{\begin{equation}}
\def\ee{\end{equation}}

\def\deg{{^\circ}}

\begin{document}
   \title{Stellar physical parameters from Str\"{o}mgren photometry} 
   \subtitle{Application to the young stars in the Galactic anticenter survey
\thanks{Catalog of the physical parameters is only available in electronic form
at the CDS via anonymous ftp to cdsarc.u-strasbg.fr (130.79.128.5) or via http://cdsweb.u-strasbg.fr/cgi-bin/qcat?J/A+A/
}}
 \author{M. Mongui\'{o} \inst{\ref{inst1}}
      \and F. Figueras \inst{\ref{inst1}}
      \and P. Grosb\o{}l \inst{\ref{inst2}}
          }
   \offprints{M. Mongui\'{o},
   \email{mmonguio@am.ub.es}}
   \institute{Departament d'Astronomia i Meteorologia and IEEC-ICC-UB,
     Universitat de Barcelona,
     Mart\'i i Franqu\`es, 1, E-08028 Barcelona, Spain \label{inst1} \and
 European Southern Observatory, Karl-Schwarzschild-Str. 2, D-85748 Garching,
Germany \label{inst2}}
   \date{Received  / Accepted }
   \abstract
{}
%Aims
{
The aim is to derive accurate stellar distances and extinctions for young stars of our survey in the Galactic anticenter direction using the Str\"omgren photometric system.  
This will allow a detailed mapping of the stellar density and absorption toward the Perseus arm.
}
%Methods. 
{We developed a new method for deriving physical parameters from Str\"omgren photometry and also implemented and tested it. This
is a model-based method that uses the most recent available stellar atmospheric models and evolutionary tracks to interpolate in a 3D grid of
the unreddened indexes $[m_1]$, $[c_1]$ and $H\beta$.
Distances derived from both this method and the classical pre-Hipparcos calibrations were
tested against Hipparcos parallaxes and found to be accurate.}
%Results.
{Systematic trends in stellar photometric distances derived from empirical calibrations were detected and quantified.
Furthermore, a shift in the atmospheric %$H\beta$ index of the %Castelli and Kurucz (2006) atmosphere 
grids in the range $T_{eff}$=$[7000,9000]$\,K was detected and a correction is proposed.
The two methods were used to compute distances and reddening for $\sim$12000 OBA-type stars in our Str\"omgren
anticenter survey. 
Data from the IPHAS and 2MASS catalogs were used to complement the detection of emission
line stars and to break the degeneracy between early and late photometric regions. 
We note that photometric distances can differ by more than 20\%, those derived from the empirical calibrations being smaller
than those derived with the new method, which agree better with the Hipparcos data.}
{}


\maketitle

\section{Introduction}\label{intro}

The spiral-arm structure of the Galaxy is an important factor for studying the morphology and dynamics of the Milky Way. 
We still lack a complete theory to describe its nature, origin and evolution, however. The Perseus spiral arm, the nearest spiral arm outside the solar radius, 
has been studied through HI neutral gas \citep{1967IAUS...31..143L}, large-scale CO surveys \citep{2001ApJ...547..792D}, star-forming complexes \citep{2003A&A...397..133R},
or open clusters \citep{2008ApJ...672..930V}, among others. 
These studies traced the arm in the second and third quadrant, but there are very few analysis linking the two quadrants, and pointing toward $\ell\sim180\degr$. 
Our project aims to fill this gap, that is, to trace the Perseus arm in the anticenter direction, both in terms of density and kinematics. 
We examine: 1) the stellar overdensity associated with the spiral arm, 
2) the position of the dust layer relative to the arm,  and 
3) the radial velocity perturbation associated with the arm which will shed light on the mechanisms driving them
\citep[e.g.,][]{2011MNRAS.410.1637S,2013MNRAS.432.2878R}.

We selected intermediate young stars (B4-A3) for this project. 
These stars are bright enough to reach large distances in the anticenter direction and old enough to 
have had time to respond to the spiral-arm perturbation. 
From stellar evolutionary models \citep{2008A&A...484..815B} we know that these stars have main-sequence 
lifetimes in the range 60-700 Myr.  Test-particle simulations evolving in a realistic Milky Way 
 spiral-arm potential demonstrated \citep[e.g.,][]{2011MNRAS.418.1423A}
that a young population -with ages even younger than 400 Myr- have had enough time 
to develop a clear stellar response to the potential spiral perturbation. 
This response is observed both in terms of stellar overdensity -redistributing the disk mass and 
tracing the spiral structure- and in terms of the induced kinematic perturbation, for instance, driving secular 
changes in the orbits of stars or generating kinematic substructure (such as moving groups). 
Furthermore, as will be discussed in our next paper (Mongui\'{o} et al. in preparation), 
this population is also not old enough to have been significantly
     affected by disk heating and therefore has a relative low
     intrinsic velocity dispersion.  Because of this lower dispersion, this
     population will respond more strongly to spiral perturbations which allows detecting
     weaker spiral amplitudes.

Str\"omgren photometry is well suited to select these stars and obtain the necessary stellar physical parameters (SPP hereafter).  
For that purpose, we conducted a deep photometric survey in the anticenter direction using the Wide Field Camera at the Isaac Newton Telescope. We obtained full 
$uvbyH\beta$ photometry for 35974 stars \citep[see][Paper I hereafter]{2013A&A...549A..78M}.

In the current paper we focus on deriving the SPP for young stars from Str\"omgren photometry. 
Our main goal is deriving accurate distances and reddening for faint stars (up to $V\sim 18^{\unit{m}}$). 
As discussed in \cite{2004AJ....127.1227C}, the empirical calibrations available up to now are derived from a small number of stars located in the solar neighborhood 
(usually clusters), which means that extrapolating these calibrations to large distances is not straightforward.  
We reviewed the classical methods based on pre-Hipparcos empirical calibrations (EC method hereafter, e.g., \cite{1978AJ.....83...48C,1979AJ.....84.1858C} or references in \cite{1991A&AS...87..319F}, among others)
by comparing photometric distances derived from them with Hipparcos parallaxes.  
But, we here proceed with a completely different approach. We developed a new model-based  method (MB hereafter) using theoretical 
atmospheric grids and stellar evolutionary models. 
At present, the Str\"omgren photometric grids available from model atmospheres \citep{1997A&A...328..349S,2004astro.ph..5087C,2006A&A...454..333C} 
cover a broad range of the stellar parameter space, which makes this approach feasible today.

In Sect. \ref{ECMB}, we describe EC and MB methods.
In Sect. \ref{checkgrid} they are applied to Hipparcos data with the aim of
checking and improving the calibrations. Biases and trends are discussed in this section. 
A rigorous statistical treatment requires  estimating individual errors for all the SPP  to reach a good and precise error estimate in the photometric 
distances and reddening. We present a Monte Carlo method that accounts for this. 
In Sect. \ref{Aplcat} the SPP for the OBA-stars in the Galactic anticenter survey are computed, and the results for the two methods are compared. 
External data provided by 2MASS and IPHAS add information to the survey, which is useful for detecting emission line stars and for defining 
reliability indexes that list inconsistencies
between different sets of data. 
Finally, we summarize the main conclusions and results in Sect. \ref{Concl}.

\section{Methods for computing the SPP}\label{ECMB}
\subsection{Classical method: based on empirical calibrations}\label{Fig91}
Several  empirical calibration methods
are available for computing the SPP from the Str\"{o}mgren photometric indexes. 
These procedures follow two steps: 1) the classification of the stars in different photometric regions, and
2) the use of empirical calibrations to obtain the intrinsic indexes and the absolute magnitudes from which 
interstellar extinction and distances can be computed.
From these, $T_{eff}$ and $\log g$ have usually been obtained using the atmospheric grids
 \citep[e.g.,][]{1985MNRAS.217..305M}, whereas ages and masses have been obtained from stellar evolutionary models 
 \citep[e.g.,][]{1997A&A...322..147A}. 
Here we evaluated these calibrations that provide distances and interstellar extinction, 
which are necessary parameters  for detecting the Perseus spiral arm.

\subsubsection{Classification methods} \label{sectclas}
\begin{figure}\centering
 \resizebox{\hsize}{!}{\includegraphics{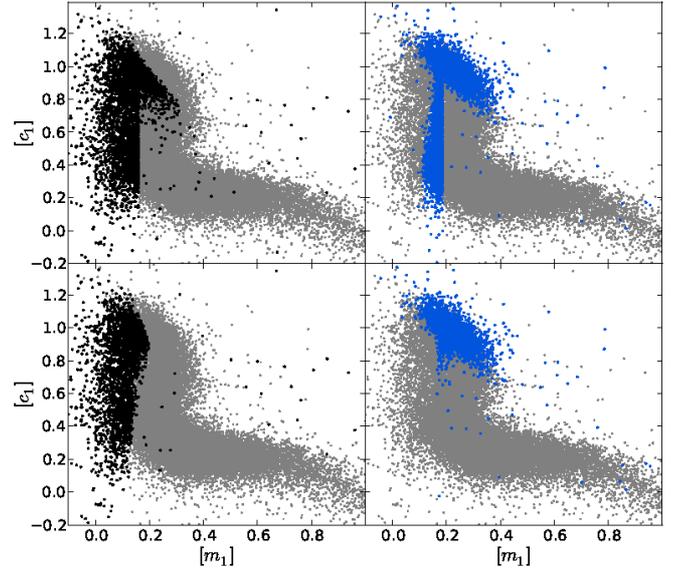}}
\caption{$[m_1]-[c_1]$ diagram for all the stars in the anticenter survey. The stars are plotted in gray and the overplotted colors show different regions: 
the early region in black, the intermediate region in blue. The classifications applied are those from FTJ91 (top-left), LI80 (top-right) and NC (bottom).}
\label{CompClas}
\end{figure}

\begin{figure}\centering
 \resizebox{\hsize}{!}{\includegraphics{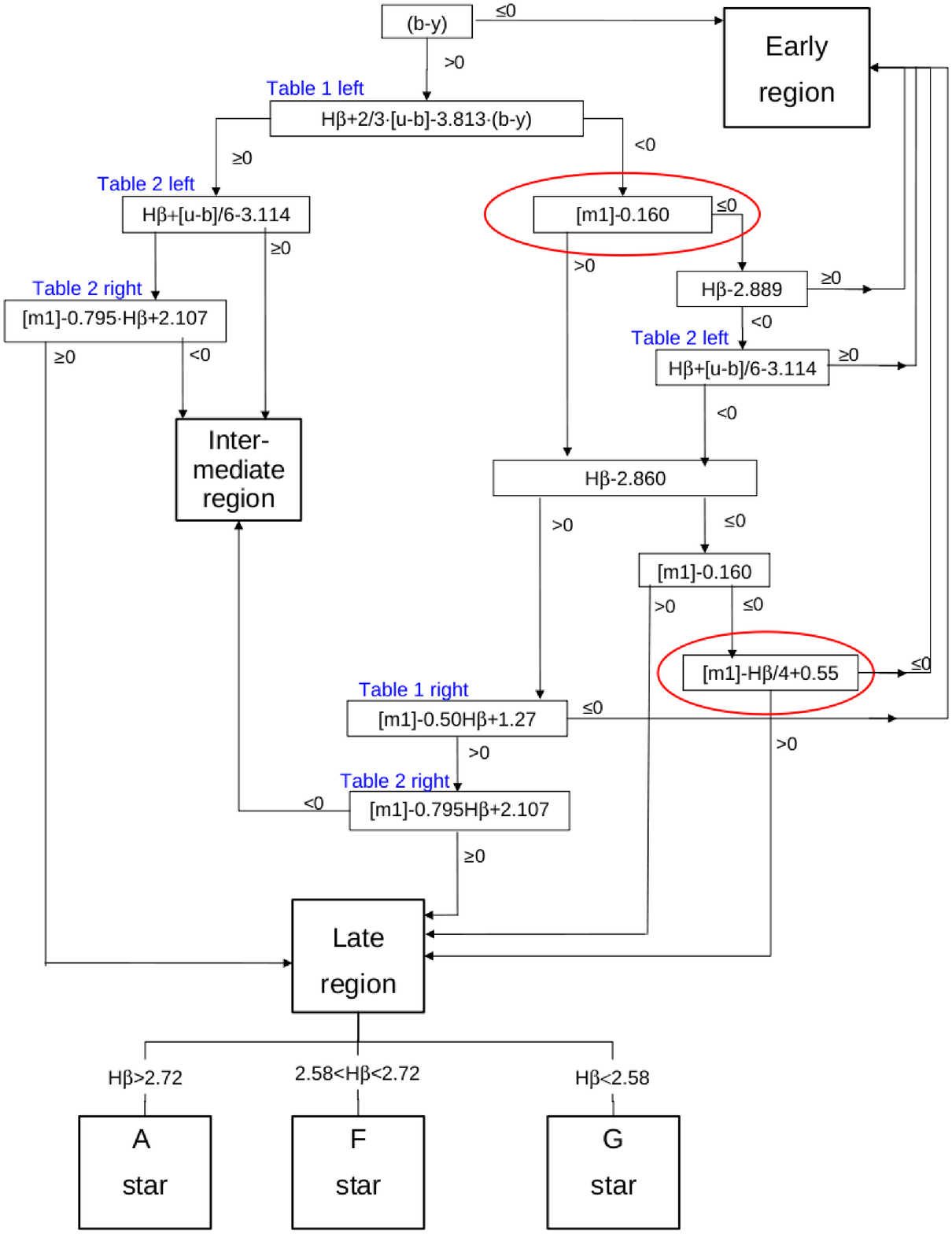}}
\caption{Scheme of the new classification method. Red circles indicate new criteria added to previous classification methods.
In blue, reference tables from \cite{1966ARA&A...4..433S}.}
\label{NC}
\end{figure}

\cite{1966ARA&A...4..433S} classified the stars in different photometric regions according to their spectral type:
the early region (B0-B9), the intermediate region (A0-A3), and
the late region, subdivided into three subregions, A3-F0, F0-G2, and later than G2.
Later on, \citet[][LI80]{1980StoOR..17.....L} and \citet[][FTJ91]{1991A&AS...87..319F} published minor changes with the
main differences being in the gap between the early and late regions in the $[m_1]-[c_1]$ plot\footnote{$[m_1]=m_1+0.33 \cdot (b-y)$ and $[c_1]=c_1-0.19\cdot (b-y)$} of 
 reddening-free indexes \citep{1966ARA&A...4..433S,1976PASP...88..917C}.
Figure \ref{CompClas} shows the $[m_1]-[c_1]$ plot for the stars in our anticenter survey, which reach magnitudes up to $V\sim18^{\unit{m}}$.  
While these methods provide almost equivalent assignments
when the stars in the early and late regions are clearly separated (i.e., for small photometric errors),
two important drawbacks have been detected
for faint stars with substantial photometric errors (see Figs. 3 and 4 from Paper I). 
First, LI80 erroneously assigned stars in the gap, i.e., for low $[c_1]$ values, to the intermediate region (see Fig. \ref{CompClas} top-right). 
Second, FTJ91  erroneously classified  stars with $[m_1]>0.16$ and high $[c_1]$ values, 
that clearly belong to the intermediate or late regions, as early-region stars (see Fig. \ref{CompClas} top-left). 
To remedy that, two additional conditions are included in the classification scheme  (indicated in red in Fig. \ref{NC}). 
The application of this new classification method (hereafter NC) to our anticenter survey is shown in Fig. \ref{CompClas} bottom,
 where we can observe that the two errors have been corrected.

\subsubsection{Empirical calibrations}

Several  calibrations  have been published in the past fifty years. An exhaustive pre-Hipparcos compilation was
published by \citet{1991A&AS...87..319F}. 
Later on, some post-Hipparcos calibrations have been established, but all of them concerning F-G late-type stars \citep[e.g.,][]{2007A&A...475..519H, 2010NewA...15..444K}.
In the present work, to derive the SPP of stars earlier than A9 we reviewed the following most often used calibrations:
\citet[][CR78]{1978AJ.....83...48C}, \citet[][LI81]{1981StoOR..18.....L} and \citet[][BS84]{1984MNRAS.211..375B} for the early region; 
\citet[][CL74]{1974edfe.book.....C}, \citet[][GR78]{1978A&AS...32..409G}, \citet[][HI83]{1983MNRAS.204..241H}, 
and \citet[][MO85]{1985MNRAS.217..305M} for the intermediate region; and  \citet[][CR79]{1979AJ.....84.1858C} for the late region up to A9.

\subsection{New strategy: the model-based method}\label{ATLAS9}

In this section we propose a new method  based on the most upgraded theoretical atmospheric grids and evolutionary models. This new approach is, by
construction, completely independent of the previously used classical method.
As input data, it simultaneously uses the three extinction-free photometric indexes $[m_1]$, $[c_1]$, and $H\beta$.
First, after an a priori selection of the metallicity of the star, a 3D fit in the theoretical $[m_1]-[c_1]-H\beta$ plane derived from the
atmospheric grid allows us to determine $\log g$ and $T_{eff}$, the intrinsic colors, and the bolometric correction $\left(BC=M_{bol}-M_V\right)$.
The absorption $\left(A_V\right)$ is then easily estimated from the observed color index.
As a second step, an interpolation in the stellar evolutionary tracks provides the luminosity $\left(L/L_{\odot}\right)$, the
mass, and the age of the star. For this, we used the interpolation code developed by \citet{1997A&A...322..147A}.
This luminosity is then translated into bolometric magnitude following $M_{bol}=4.74-2.5\log\left(\frac{L}{L_{\odot}}\right)$.
Finally, the distance is computed using the $BC$, the $A_V$, and the apparent $V$ magnitude.

The most recent atmospheric grids are those from \citet{2004astro.ph..5087C,2006A&A...454..333C} using ATLAS9 and the mixing length 
theory\footnote{http://wwwuser.oat.ts.astro.it/castelli/colors.html}. For each grid point, the authors provide $T_{eff}$, $\log g$, unreddened indexes ($(b-y), m_1, c_1$ and $H\beta$)
and bolometric correction ($BC$).
A different set of grids was previously published by \citet[][hereafter SK97]{1997A&A...328..349S}, using in this case turbulent convection models for late-type stars from
\citet{1991ApJ...370..295C,1992ApJ...389..724C}, which have significant differences from the first one.
Unfortunately, SK97 do not provide the $H\beta$ values, therefore a direct comparison of the results has only been
possible by combining SK97 data with the $H\beta$ values from \citet{2006A&A...454..333C} or from \citet{1995A&A...293..446S}, who also used  \citet{1979ApJS...40....1K} grids
and the mixing length theory. When we compared the two sets we found differences of up to 0$\fm$025 in $(b-y)_0$ and 0$\fm$06 in $c_1$  for cold stars. 
Differences for hot stars were always smaller that 0$\fm$02. Differences in $m_1$ are small for low gravities and low
$T_{eff}$, but then they increase up to 0$\fm$03 for $\log g$=4.5-5.
The differences between different grids are revisited in Sect.\ref{compMB}.

Before this method is applied, it is important to analyze what it requires by way of an a priori knowledge of the star's metallicity. 
This parameter is needed to select the atmospheric grid and the evolutionary model to be used. 
We checked that the degeneracy in the $[m_1]-[c_1]-H\beta$ for several metallicities is too high to derive this parameter simultaneously with the other SPP. 
Thus, this possible drawback has to be properly analyzed. We identified two options. First, for the stars A3-A9 in the late region it is possible 
to derive $[Fe/H]$ from the photometric indexes using empirical calibrations \citep[see][and the references therein]{1989A&A...221...65S}, 
which means that in this case one could use its corresponding 
interpolated grids and evolutionary models. Second, one can consider different a priori values for the metallicity of the star (or  the working sample) and quantify 
whether the changes in SPP values are significant. 
We considered this second option when applying the MB method to our anticenter survey from Paper I. 
These stars have distances of about 3-4 kpc toward the Galactic anticenter.  Assuming a radial metallicity gradient of about -0.1 dex/kpc -indeed a very 
uncertain parameter at present- the metallicity of the stars in the sample will mostly be in the range [0,-0.5] dex. SPP data were computed using the 
Castelli grids available for $[Fe/H]$=-0.5 and $[Fe/H]$=0.0. We checked that for hot stars ($T_{eff}>7000$K), the differences in
$[m_1]$ and $[c_1]$ are smaller than 0$\fm$02 and the differences in $H\beta$
smaller than 0$\fm$01, that is, they are on the same order of the photometric errors. 
On the other hand, we mention that the EC method is not free from this drawback. Although empirical calibrations for deriving
$[Fe/H]$ are available in the literature, the intrinsic indexes and absolute magnitude are usually calibrated from
a solar neighborhood sample (see Sect. \ref{Fig91}), therefore equivalent trends and biases can be expected when this is applied to the most distant stars.

The 3D fitting algorithm we developed (see Appendix \ref{Ap3Dfit}) maximizes the probability for one star to belong to a point of the $[m_1],[c_1],H\beta$ 
theoretical grid. It takes into account the distance between them as well as the photometric errors in the three indexes, which are assumed to be Gaussian.
We took into account that stars in the gap between the early and late regions can have a two-peak probability distribution, that is, 
they can belong to either one or the other side of the gap.
 To keep track of this, both the first ($P_{max}$) and secondary ($P^{(B)}_{max}$) maximum probabilities are stored together with their corresponding SPP.
When $P_{max}/P^{(B)}_{max}$ is close to one, that is, when a star has similar probabilities to belong to the early or late regions, these probabilities allow us to 
detect the ambiguity and to point out the need for additional information (e.g., 2MASS; see Sect. \ref{Sect2M}) for the final assignment of the 
SPP.

After $T_{eff}$, $\log g$, and $BC$ were derived, the stellar evolutionary models of \citet{2008A&A...484..815B,2009A&A...508..355B} were used to derive the visual absolute magnitude, and in turn the stellar
distance. For a fixed metallicity, \citet{2008A&A...484..815B,2009A&A...508..355B} provided 32 evolutionary tracks with masses between 0.6 and 20 $M_{\odot}$.
For more massive stars, that is, for stars earlier than $\sim$O8 (not used in our analysis of the spiral-arm overdensity),
the \citet{1993A&AS..100..647B} evolutionary tracks of the same Padova database were implemented.
These tracks cover the 20-120$M_{\odot}$ range that is not reached by \citet{2008A&A...484..815B,2009A&A...508..355B} grids.
We used the interpolation code provided by \citet{1997A&A...322..147A}.

\subsection{Errors}\label{errors}
The errors for the obtained SPP were estimated through Monte Carlo simulations. For each star, 100 realizations were performed and the corresponding
observed parameters were sampled ($V$, $(b-y)$, $m_1$, $c_1$, and $H\beta$) assuming random Gaussian errors. 
The mean and dispersion for each SPP parameter were obtained from these 100 realizations. This procedure was applied to the EC and MB methods. 
For the EC method one could also follow the strategy proposed by \citet[][for B-type stars]{2008A&A...486..471R}  and \citet[][for A-type stars]{1978A&AS...33..347K}, 
which is based on the error propagation in the empirical relations. 
We note several advantages of the Monte Carlo method proposed here: 
1) it provides a better consistency between the errors
computed from EC and MB, an important fact when trying to test these two methods against Hipparcos data (Sect. \ref{checkgrid}), 
2) this error computation process applied to the EC method also provides a parameter related
to the probability of the star to be assigned to a region different from the one initially assigned ($N_{reg}$), 
3) equivalent parameters are obtained for the MB method, that is,
the 3D fitting algorithm can assign some of the realizations to the other side of the gap. The parameter $N_{side}$ indicates the
number of realizations located at the same side of the gap and provides an additional flag that accounts for a good assignment.

A large portion of the stars in our samples may be binaries, which can include a bias in the distances when we treat them as single stars. 
The effects of this bias are discussed in Appendix \ref{BinMet}, where some simulations were made to understand the variation in the photometric indexes caused by
 a secondary star. The larger biases are for mass ratios close to $M_2/M_1\sim1$, giving upper limits of 0$\fm$05 for $[c_1]$, 
0$\fm$02 for $[m_1]$, and 0$\fm$04 for $H\beta$ for stars with $T_{eff}>$7000\,K. 
In that case, the error in apparent magnitude can reach 0$\fm$75, that is, a 30\% error in distance.

\section{Testing photometric distances using Hipparcos data}\label{checkgrid}

Hipparcos data were used here to determine whether  the EC and MB methods can derive good stellar distances. 
Because we are dealing with nearby stars, 
interstellar extinction plays a secondary role and calibrations used to derive $A_V$ cannot be tested.  We need to detect, evaluate, and correct for,
when possible, the systematic trends observed when comparing trigonometric and photometric distances. Hipparcos OBA-type stars with Str\"omgren 
indexes  from \cite{1998A&AS..129..431H}  were taken from the compilation of \cite{2000A&A...359...82T}  -for the OB stars- and \cite{1998PhDT........21A} -for A type stars. 
The final sample contains 4601 OBA with parallaxes and all photometric indexes (including $H\beta$):  91 O-stars, 2568 B-stars, 1574 A0-A3 stars, and 368 A4-A9. The spectral type 
was taken from the Hipparcos Input Catalog (HIC hereafter).

\begin{figure}\centering
 \resizebox{9cm}{!}{\includegraphics{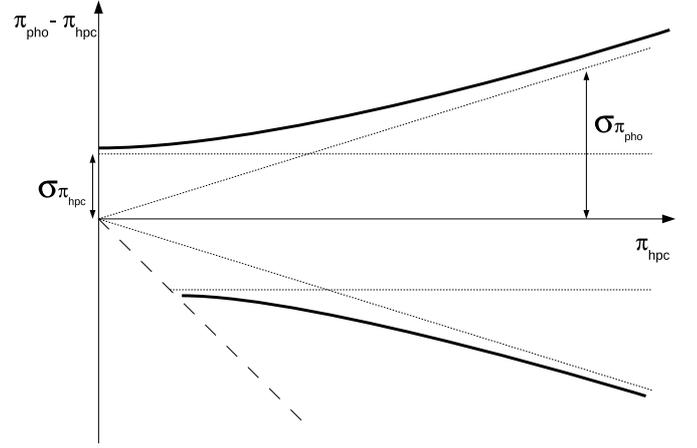}}
\caption{Scheme of the expected distribution of $(\pi _{pho}-\pi_{hpc})$ vs. $\pi_{hpc}$. 
Here we assume a constant absolute astrometric standard error in the trigonometric Hipparcos-based 
parallax $\sigma_{\pi_{hpc}}$. 
As can be seen in Fig.17.7 of Hipparcos Catalog % (Vol.3, ESA SP-1200, pag. 381)
this assumption is correct only for bright stars. 
We checked that the 85\% of the stars considered in this analysis, with the requirement to have full  
Str\"omgren photometry have $V<8^{\unit{m}}$, therefore the assumption is acceptable. 
For photometric parallaxes, a constant relative error is assumed. 
Again, in this case, as the stars are apparently bright, which means that they are nearby, we assume that
 reddening effects and a low signal-to-noise ratio do not affect the derived distances. }

\label{expar}
\end{figure}

To undertake this analysis it is important to work in the space of the observables. To do this, we compared the methods in terms of parallaxes ($\pi$), and not in terms of 
distances ($r$). As is known, whereas the error in the Hipparcos trigonometric parallax is symmetric, the corresponding error in distance is not \citep{1997ESASP.402..449L}. 
Our strategy was to check the difference between the photometric and the trigonometric parallax ($\pi_{pho}-\pi_{hpc}$) against trigonometric parallaxes ($\pi_{hpc}$).  
We preferred the trigonometric parallax in the abscissa axis because it has a constant absolute error, which is a more clear and understandable behavior.  
Figure \ref{expar} shows the scheme of the considerations that were taken into account in this comparison.  
First, whereas the absolute error in trigonometric parallax is constant, that is, it does not depend on parallax, photometric parallaxes have, by construction, 
a constant relative error, in other words, an absolute error that linearly depends on parallax.  The propagated error on the difference ($\pi_{pho}-\pi_{hpc}$) is the convolution of  both 
(see black curve in Fig. \ref{expar}). On the other hand, we know that while about 3\% of our sample have negative trigonometric parallaxes (removed from the sample), 
photometric parallaxes are always, again by construction, defined positive. Although statistically some negative values for small photometric parallaxes 
(large distances) are expected, we forced all the photometric parallaxes to be positive. This creates a forbidden region below the line of 45$\degr$ (shown as a dashed line in Fig. \ref{expar})
because the differences  ($\pi_{pho}-\pi_{hpc}$)  will be always larger than  ($-\pi_{hpc}$). 
The effects caused by this constraint were taken into account by avoiding the smallest parallaxes -the entire analysis was made considering $\pi_{hpc}<$3\,mas- 
and by working with medians instead of means. The first condition allowed us to remove stars with significant interstellar extinction.  
The analysis was made separately for each of the Str\"omgren regions (early, intermediate, and late) separately.  
To have a significant number of stars per bin when computing medians, we constrained the analysis to stars with  $\pi_{hpc}<$10\,mas for the early region (O-B9), 
$\pi_{hpc}<$15\,mas for the intermediate, and $\pi_{hpc}<$20\,mas for the late group (A3-A9).  Results are presented in Figs. \ref{HPCEC} and \ref{HPCMB}. 

\subsection{Checking the classical EC method }
\begin{figure}\centering
 \resizebox{\hsize}{!}{\includegraphics{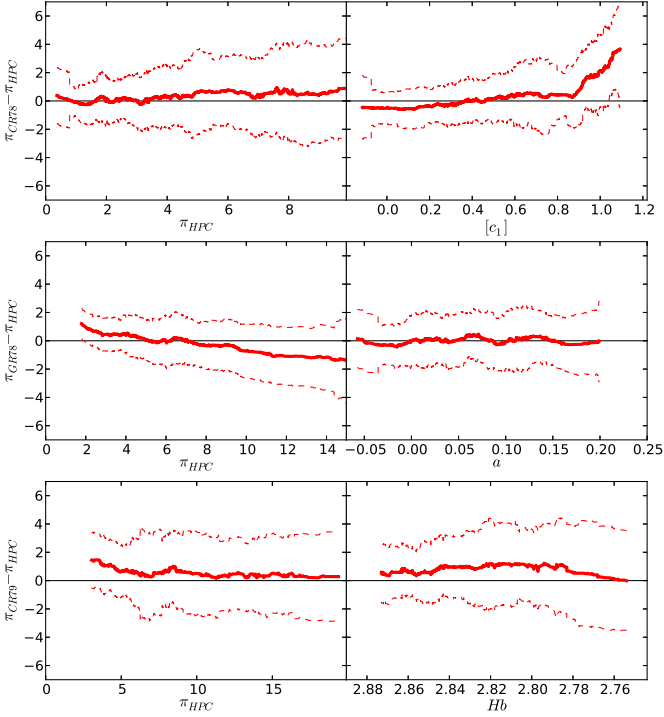}}
\caption{Running median for the difference between EC photometric and  Hipparcos parallaxes vs. Hipparcos parallax (left) and vs. $T_{eff}$ photometric index indicator. 
Top: early region using CR78 ($[c_1]$ indicating $T_{eff}$). Middle: intermediate region using GR78 ($a$ indicating $T_{eff}$). Bottom: late region up to A9 using CR79 
($H\beta$ indicating $T_{eff}$). The solid line shows the moving median and the dashed line is one standard deviation. Parallaxes are given in mas.}
\label{HPCEC}
\end{figure}

In  Table \ref{Dparmed} we detail the median differences between Hipparcos and photometric parallaxes, as well as the trends in $T_{eff}$, for the different photometric regions.
For the early region, CR78 and BS84 give very similar results, with the median close to zero. 
A clear trend in $[c_1]$ -tracing $T_{eff}$ for this region- is present, however, with $r_{hpc}<r_{phot}$ for hotter stars and $r_{hpc}>r_{phot}$ for the colder B-type stars. 
 For the intermediate region,
we found very similar results for all the methods, with CL74 and GR78 with a lower error in median, and always obtaining $r_{hpc}>r_{phot}$. 
A weak trend in distance is present for all of them, with no clear trend 
in $a=(b-y)+0.18 \cdot \left((u-b)-1.36\right)$ - tracing $T_{eff}$ in this region.
For the late region up to A9, the 
CR79 method slightly underestimates the distances with a mild trend in $H\beta$ -the $T_{eff}$ indicator in this region. 
The calibrations used in the following sections for the EC method give smaller biases, 
that is, CR78 for the early region, GR78 for the intermediate region, and CR79 for the late region up to A9. The results for
these three methods are shown in Fig. \ref{HPCEC}.

\begin{table*}
 \caption{Median differences between Hipparcos and photometric parallaxes in mas. Each region was selected according to both HIC and NC classifications in the same case (most restrictive case). 
The median for several intervals traces the trend in $T_{eff}$.}
\label{Dparmed}
\centering
\begin{tabular} {c|ccccc}
early&All&$[c_1]<$0.2    & 0.2$<[c_1]<$0.5&0.5$<[c_1]<$0.8&0.8$>[c_1]$    \\\hline
$\langle \pi_{CR78}-\pi_{hpc}\rangle$&0.07$\pm$0.08&-1.97$\pm$0.22&-0.51$\pm$0.11&0.18$\pm$0.12&1.32$\pm$0.20\\
$\langle \pi_{LI80}-\pi_{hpc}\rangle$&0.23$\pm$0.09&-1.97$\pm$0.22&-0.51$\pm$0.11&0.32$\pm$0.12&2.05$\pm$0.19 \\
$\langle\pi_{BS84}-\pi_{hpc}\rangle$ &-0.16$\pm$0.07&-1.75$\pm$0.25&-0.60$\pm$0.10&-0.06$\pm$0.10&0.70$\pm$0.13\\\hline
 interm. & All   & $a<$0.05    &0.05$<a<$0.1&0.1$<a$   &\\\hline  
$\left\langle\pi_{CL74}-\pi_{hpc}\right\rangle$&0.45$\pm$0.09&0.18$\pm$0.16&0.47$\pm$0.13&0.61$\pm$0.20\\
$\left\langle\pi_{GR78}-\pi_{hpc}\right\rangle$&0.48$\pm$0.09&0.17$\pm$0.16&0.53$\pm$0.13&0.70$\pm$0.20\\
$\left\langle\pi_{HI83}-\pi_{hpc}\right\rangle$&0.53$\pm$0.10&0.37$\pm$0.17&0.57$\pm$0.14&0.65$\pm$0.22\\
$\left\langle\pi_{MO85}-\pi_{hpc}\right\rangle$&0.58$\pm$0.10&0.36$\pm$0.17&0.58$\pm$0.15&0.74$\pm$0.24\\\hline
late &All &              2.85$<H\beta$     &2.8$<H\beta <$2.85&$H\beta <$2.8     \\\hline
$\left\langle\pi_{CR79}-\pi_{hpc}\right\rangle$&0.62$\pm$0.18 &0.19$\pm$0.34&0.97$\pm$0.23&0.20$\pm$0.38\\
\end{tabular}

\end{table*}

\subsection{Checking the new MB method: Correcting for the atmospheric model}\label{compMB}
\begin{figure}\centering
 \resizebox{\hsize}{!}{\includegraphics{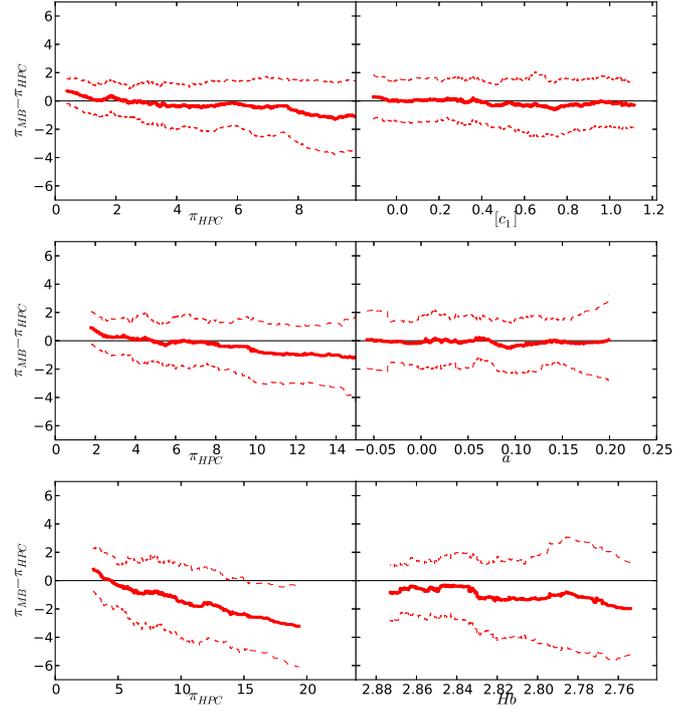}}
\caption{Running median for the difference in parallax vs. Hipparcos parallax (left) and vs. $T_{eff}$ indicator for each region using the MB method. 
Top: early region ($[c_1]$ indicating $T_{eff}$). Center: intermediate region ($a$ indicating $T_{eff}$). Bottom: late region up to A9 ($H\beta$ indicating $T_{eff}$). All of them use Castelli grids.
The solid line shows the moving median and the dashed line is one standard deviation. Parallaxes are given in mas.}
\label{HPCMB}
\end{figure}

For the MB method we also compared the results for the three photometric regions. 
For the early region, results give a small bias in parallax of -0.51\,mas, 
but in this case, the trend in effective temperature shown by CR78 is removed (see Fig. \ref{HPCMB} top). 
For the intermediate region, the bias is small (0.03\,mas in median), giving better results than the EC method
(see Fig. \ref{HPCMB} middle). 
For the late region we found a clear bias between MB and Hipparcos parallaxes with a
trend in both parallax and $T_{eff}$ (see Fig. \ref{HPCMB} bottom). We have checked that there is also a bias in $E(b-y)$, with clearly negative mean
values (-0$\fm$03). 
The check was repeated using the atmospheric grids  by \cite{1995A&A...293..446S} for $uvby$, with different combinations of the grids by
\cite{2006A&A...454..333C} and \citet{1997A&A...328..349S} for the  $H\beta$ values, following
\begin{equation}
H\beta_k = k\cdot H\beta_{SK97}+(1-k)\cdot H\beta_{CK06}.
\end{equation}
These options modify the results for stars with 5500$<T_{eff}<$8500\,K in the correct direction.
The best results were achieved for the \cite{1995A&A...293..446S} grids for $uvby$ with a combination of the grids of \cite{2006A&A...454..333C} and \citet{1997A&A...328..349S} for the $H\beta$,
with a value of $k=$0.2.

\begin{figure}\centering
 \resizebox{\hsize}{!}{\includegraphics{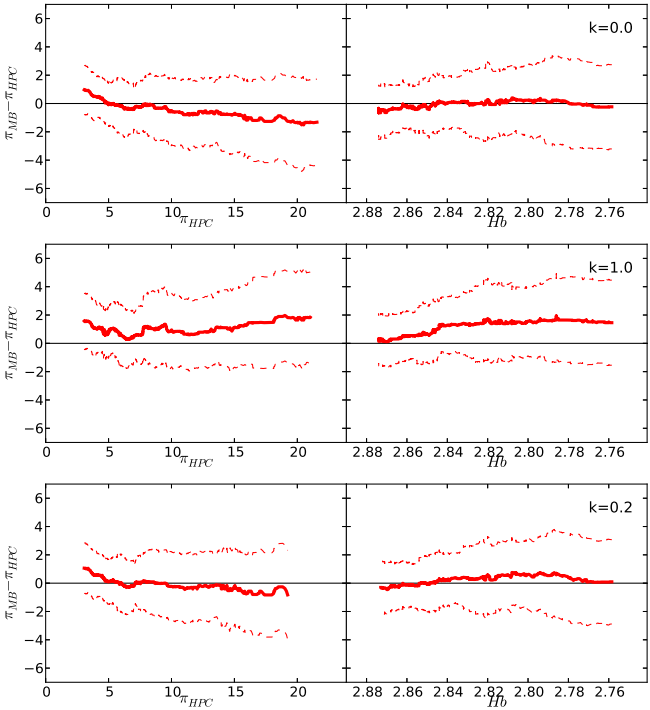}}
\caption{Running median for the difference between MB  and Hipparcos parallaxes vs. Hipparcos parallax (left) and vs. $T_{eff}$ indicator for the late region (A3-A9). 
Top: $k$=0. Center:  $k=1$. 
Bottom: $k$=0.2. The solid line shows the moving median and the dashed line is one standard deviation. Parallaxes are given in mas.}
\label{HPCMBcor}
\end{figure}

From this we suggest that there is a bias in the Castelli grids for late type stars. This problem was solved by creating a new grid as a combination of other available options. 
The new solution shows no bias, as can be seen in Fig. \ref{HPCMBcor}. 
The new grids were checked only for stars with $T_{eff}>$7000\,K.


\section{SPP for young stars in our anticenter survey}\label{Aplcat}
We applied the EC and MB method to the young stars in our Galactic anticenter survey (Paper I). 
The full catalog contains
35974 stars with all Str\"omgren indexes, the extended catalog lists 96980 stars with partial data.  
The inner 8$\deg^2$ and the outer 8$\deg^2$
reach $\sim$90\% completeness at V$\sim17^{\unit{m}}$ 
and $V\sim 15\fm$5, respectively (see Paper I).
Photometric internal precisions between 0$\fm$01-0$\fm$02  were obtained for stars brighter than  $ V=16^{\unit{m}}$  with several measurements,
increasing up to 0$\fm$05 for  fainter stars  with $V\sim$18$^{\unit{m}}$ (see Fig. 3 in Paper I).
During the process of computing the SPP with EC and MB,
a discrepancy between the distribution of $H\beta$ values and expected range was detected.
This led us to re-check the list of standard stars used\footnote{An emission line star (S3R1N4), 
a possible T-Tauri variable star (S4R2N12), stars with an inconsistent $H\beta$ value (S3R2N8 and S3R2N7) 
and stars with large discrepancies between differences sources of information (S2R2N46, S2R1N25 and S3R1N13) 
were rejected from the list of standards. ID labels correspond to \cite{2001AJ....121.2075M}.}
 and the equations for the transformation into the standard system.

In addition, we realized that the equation without $(b-y)$ term (i.e., Eq. 3b in Paper I) 
provides more coherent results.
Both changes lead to slightly different new values for the $H\beta$ indexes, presented in this 
new version of the catalog. The current coefficients for the transformation into the standard system are presented in Appendix \ref{trans2}.

The EC method was applied to 11854 stars classified as belonging to the early, intermediate, or late regions ($<$A9) using the NC classification method.
This provides intrinsic color, $A_V$, $M_V$, and distances, as well as the corresponding errors.
In addition, the  catalog provides the photometric region assigned ($reg_{NC}$) and the parameters  $N_{reg}$ and $N_{regi}$, which are
 reliability indicators of this assignment (see Sect. \ref{errors}).  

\begin{figure}\centering
  \resizebox{\hsize}{!}{\includegraphics{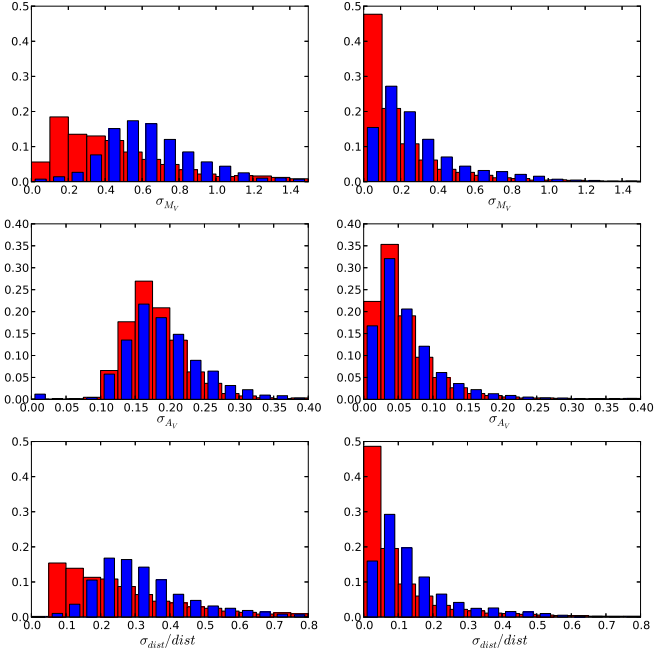}}
\caption{Error distribution for $M_V$, $A_V$ and distances obtained with the MB method (red) and EC method (blue). 
Left: stars with only one measurement. Right: stars with more than one measurement.}
\label{FPMBN}
\end{figure}

When we applied the MB method to our survey we obtained  13337 stars with $T_{eff}>$7000\,K.
Distance, intrinsic colors, $M_V$, $T_{eff}$, 
$\log g$, age, mass, and $BC$, together with their corresponding errors, are provided. 
Additionally, parameters indicating the quality of the 3D fitting (see Sect. \ref{errors}) and the SPP values for the alternative assignment 
(due to the uncertainty between the early and late regions for faint stars, see Sect. \ref{ATLAS9}) are given.
The accuracy obtained for $M_V$, $A_V$, and distances are presented in Fig. \ref{FPMBN} and are compared with those of the EC method. 
The errors in distances and $M_V$ derived with the 
EC method are clearly larger than those derived with the MB method. 
As explained in Paper I, 
the  published error in the observed indexes are 
 the error of the mean for the stars with more than one measurement  (32\% of the stars), and
the standard deviation computed by error propagation when there is only one measurement (68\% of the stars). In the first case, the errors in distances are smaller than 20\%,
while for stars with one measurement, they can reach 40-50\% in some cases.

\subsection{Comparison between the MB and EC methods}
\begin{figure}\centering
  \resizebox{\hsize}{!}{\includegraphics{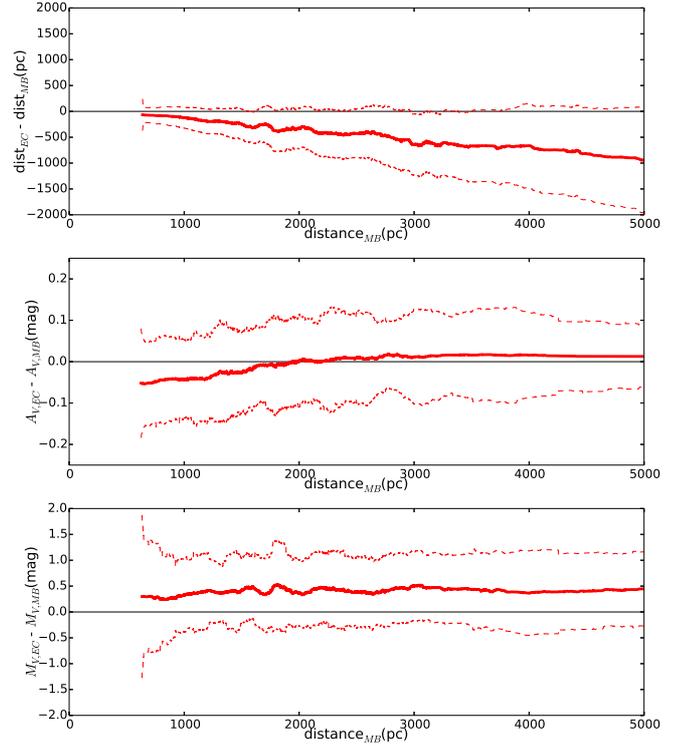}}
\caption{Running median for the differences in distance (top), $A_V$ (center) and $M_V$ (bottom) obtained from the two methods (EC minus MB) vs. distances from the MB method for the 
stars in our anticenter survey.
The solid line shows the moving median and the dashed line is one standard deviation.}
\label{compFP}
\end{figure}

Figure \ref{compFP} shows the differences in distance, $A_V$, and $M_V$ obtained between the two methods. 
From this analysis, about 8\% of faint stars of the sample were excluded, because they are placed in the  gap 
and the EC and MB methods assign them to different regions. Including them would mask the statistical comparison between the methods.
The first important trend observed in these figures is that the EC method gives clearly smaller distances than the MB method.
We can state that a systematic difference exists between the two methods, which provides differences in relative distances as large as 20\%, that is, differences as large as 
300-500\,pc can be present near the expected position of the Perseus arm, at about 2-3\,kpc. 
This shift is in the same direction as the clear trend found when comparing EC calibrations and the Hipparcos parallaxes in Table \ref{Dparmed} and Fig. \ref{HPCEC}: 
the EC provides smaller distances than Hipparcos in most of the cases. This common behavior suggests a possible bias in the 
EC pre-Hipparcos calibrations published by Crawford in the seventies.

\begin{figure}\centering
  \resizebox{\hsize}{!}{\includegraphics{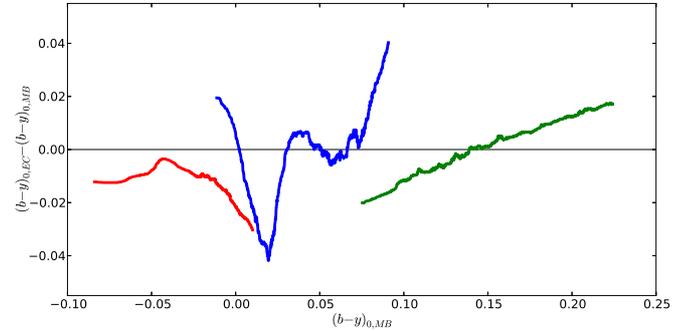}}
\caption{Running median of the differences in $(b-y)_0$ (EC - MB) vs.  $(b-y)_0$ for early (red), intermediate (blue), and late-type stars (green).}
\label{compbyo}
\end{figure}

In the $M_V$ plots (Fig. \ref{compFP} bottom) we can quantify this effect in terms of the visual absolute magnitudes. A constant bias in EC, of about 0$\fm$4-0$\fm$5,
makes the stars systematically intrinsically faint, that is, more nearby than they really are. 
The differences observed in $A_V$ are smaller (see Fig. \ref{compFP} middle) with a less significant contribution to the derivation of photometric distances. 
We observe that, for stars at large distances ($>$1.5\,kpc), the differences in $A_V$ between EC and MB are smaller than 0.02 magnitude, that is, they are on the order of the smaller photometric 
errors. On the other hand, stars at short distances ($<$1.5\,kpc) have differences as large as 0$\fm$05. 
In Fig. \ref{compbyo} we quantify the differences 
between the two methods in the intrinsic $(b-y)_0$ color. We separately plot the trends observed inside each photometric region, according to the NC classification.  
The differences are small, reaching only values up to 0.04 magnitude around A0 and A3 type stars, that is, at the edges between two photometric regions. 
We consider that this behavior reflects another disadvantage of the empirical calibration methods. By construction, the MB method should not lead to any special jump around A0 and A3, 
the edges of the Str\"omgren photometric regions. Thus, the opposite differences observed around these edges in Fig. \ref{compbyo} 
- with positive or negative differences depending on whether we approach from the left or from the right - will be caused by the EC method. 
We suggest that calibrations that are valid in the center of the photometric region are erroneously extrapolated to its edges. 

\subsection{2MASS data as a reliability test}\label{Sect2M}
\begin{figure}\centering
  \resizebox{\hsize}{!}{\includegraphics{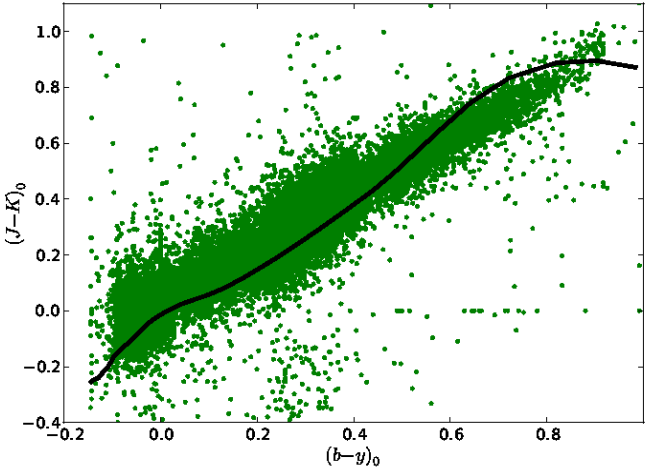}}
\caption{ $(J-K)_0$ vs. $(b-y)_0$ plot for the anticenter stars computed following the MB method. The black line shows the expected main sequence 
obtained by combining grids of the ATLAS9 model atmosphere from \citet{2004astro.ph..5087C,2006A&A...454..333C} and \cite{1998A&A...333..231B}.}
\label{2massJK}
\end{figure}

Almost all the stars in the anticenter survey have 2MASS counterparts (99.3\% of the sample). 
These data were used to detect classification problems in the MB method and to provide a quality flag to the obtained SPP data. 
The $JHK$ data are less affected by extinction. 
Using the relations from \cite{1985ApJ...288..618R}, we computed the expected absorption in the $JHK$ indexes from the 
$A_V$ derived from Str\"{o}mgren photometry. From this, intrinsic $JHK$ color indexes were computed and compared with the 
intrinsic Str\"{o}mgren index $(b-y)_0$ obtained from the MB method. 
This was made assuming the main-sequence relations (more than 90\% of the sample stars have $\log g>3$) obtained from combining the grids from
\cite{1998A&A...333..231B} for $JHK$, 
and \citet{2004astro.ph..5087C,2006A&A...454..333C} for $uvby$ (see Sect.\ref{ATLAS9}).
We computed the smallest distance in the 
 $(J-K)_0-(b-y)_0$ and  $(J-H)_0-(b-y)_0$ planes (see Fig.\ref{2massJK}) between the current location of the star and the expected main sequence, that is, $D_{JK,min}$ and  $D_{JH,min}$. 
We expect these distances to be small. 
A very high value for these $D_{JK,min}$ and  $D_{JH,min}$ indexes would indicate for most of cases that the Str\"{o}mgren photometry is not coherent with the 2MASS photometry.
These operations were repeated with the $A_V^{(B)}$ and $(b-y)_0^{(B)}$ (i.e., SPP obtained assuming that the star belongs to the 
other side of the gap), obtaining the values $D_{JK,B,min}$ and $D_{JH,B,min}$. In principle, the departures from the mean relations should be smaller in the first case,
otherwise this would indicate a misclassification problem. These differences  $Flag_{JK}=D_{JK,B,min} - D_{JK,min}$ and $Flag_{JH}=D_{JH,B,min} - D_{JH,min}$
were  added to the catalog. Seventy-six per cent of the stars have two positive indexes, which indicates that option B results have larger discrepancies,
while only around 3\% of the stars have $Flag_{JK} + Flag_{JH}<-0.1$, which indicates that the first assignment is an incorrect classification; and the SPP computed for the other side of 
the gap are preferred.

\subsection{Comparison with IPHAS data}
\begin{figure}\centering
  \resizebox{\hsize}{!}{\includegraphics{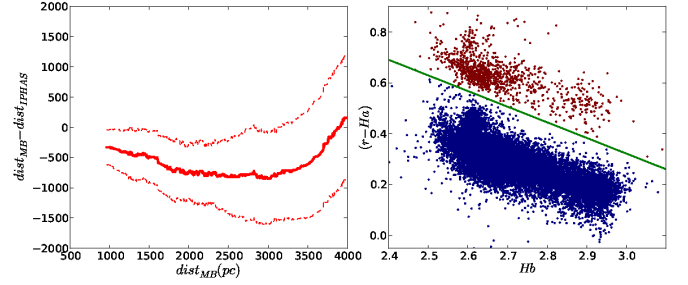}}
\caption{Left: differences between MB and IPHAS distances as a function of the MB distance for early-A type stars (A0-A5). 
The solid line shows the moving median and the dashed line is one standard deviation.
Right:  $r-H\alpha$ vs. $H\beta$. In red, stars classified as emission line stars. The green line shows the established limit. }
\label{compIPHAS}
\end{figure}

Only half of our survey area is covered by the IPHAS initial data release (IDR) \citep{2008MNRAS.388...89G}. Up to 54\% 
of the stars in our survey with full photometry have IPHAS data. 
The distance for early-A type stars can be estimated from IPHAS data following \cite{2010MNRAS.402..713S}. The authors suggested to select the early-A stars from a color-color 
diagram and  assumed for them $(r-i)$=0.06 and $M_r$=1.5. We compared these distances with those derived using MB for the A0-A5 stars in our catalog (see Fig.\ref{compIPHAS}-left).
The results give a clear bias of up to 30\%
with larger IPHAS distances. Taking into account the results from previous sections, we have $r_{EC}<r_{MB}<r_{IPHAS}$. 

IPHAS data also allow us to detect emission line stars by combining our $H\beta$ data and the $(r-H\alpha)$ index.
We consider that a star is an emission line when $(r-H\alpha)>-0.614\cdot H\beta +2.164$ (see Fig. \ref{compIPHAS}-right). 
The flag $EMLS$ is included in the catalog.
About 5\% of the stars with IPHAS data are found to be emission line stars.

\section{Summary and conclusions}\label{Concl}
We presented a new method for deriving stellar physical parameters from Str\"omgren photometry.
This method uses the three
extinction-free photometric indexes ($[c_1]-[m_1]-H\beta$) to interpolate in the theoretical atmospheric grids  deriving
$(b-y)_0$, $A_V$, $BC$, $T_{eff}$, and $\log g$. The stellar evolutionary models were then used to obtain $M_V$, distances, ages, and masses.
It rigorously takes into account the observational errors in the process, which makes it ideal for deriving physical parameters for stars collected from large and deep 
photometric surveys. Our study focused on the young and massive stars with $T_{eff} > $7000\,K ($\sim$OBA type stars).
Furthermore, 2MASS and IPHAS data were used to complement the results.

We have performed an exhaustive and accurate comparison of this new method with the classical approach, which is based on the use of pre-Hipparcos empirical calibrations, 
and with distances derived from Hipparcos parallaxes.
Substantial differences are present.
The most significant trends we found when we compared the empirical calibrations with the Hipparcos parallaxes are 
1) a trend in the photometric distance for the early region (O-B9) as a function of $[c_1]$ -temperature indicator in this spectral range.  For stars with $[c_1]<$0.2, 
there is a clear bias in the sense $r_{hpc}<r_{phot}$, while for stars with $[c_1]>$0.8 it is the opposite: $r_{hpc}>r_{phot}$.
2) a bias for the photometric distances for the intermediate and late regions (A0-A9), obtaining always lower values than for Hipparcos distances. 
  During the implementation of the MB method, a significant departure of the new distances from Hipparcos parallaxes was detected in
the range of $T_{eff}=[9000,7000]$. We proposed that this bias is
caused by a shift in the theoretical $H\beta$ index of the \cite{2006A&A...454..333C} atmospheric grids.
Hipparcos distances allowed us to quantify this correction and incorporate it into our code for a proper photometric distance derivation.

The two methods were used to obtain the stellar physical parameters for the OBA-type stars in our Str\"omgren anticenter survey (Paper
I, up to $V\sim18^{\unit{m}}$). The data are published in Appendix \ref{CAT}  using both
methods. Our final catalog contains data for more than twelve thousand OBA-type stars. 
Substantial differences of
about 20\% between the two distances are present, 
with the new method yielding the larger distances (corresponding to 0$\fm$5 in $M_V$).
In contrast, the two methods provide almost equal values for the interstellar absorption, with differences always smaller than 0$\fm$02.

In forthcoming
papers this information will be used to study the radial stellar
distribution with the aim to detect the overdensity due to the
Perseus arm. The same data will allow us to create a 3D extinction
map in our survey area, and analyze the dust distribution
and its relation to the Perseus arm dust layer.

\begin{acknowledgements}
This work was supported by the MINECO (Spanish Ministry of Economy) 
- FEDER through grant AYA2009-14648-C02-01 and CONSOLIDER CSD2007-00050.
M.Mongui\'{o} was supported by a Predoctoral fellowship from the Spanish Ministry (BES-2008-002471 through ESP2006-13855-C02-01 project).
\end{acknowledgements}

\bibliographystyle{aa}
\bibliography{Monguio}

\begin{thebibliography}{47}
\expandafter\ifx\csname natexlab\endcsname\relax\def\natexlab#1{#1}\fi

\bibitem[{{Antoja} {et~al.}(2011){Antoja}, {Figueras}, {Romero-G{\'o}mez},
  {Pichardo}, {Valenzuela}, \& {Moreno}}]{2011MNRAS.418.1423A}
{Antoja}, T., {Figueras}, F., {Romero-G{\'o}mez}, M., {et~al.} 2011, \mnras,
  418, 1423

\bibitem[{Arenou(2010)}]{LL:FA-054}
Arenou, F. 2010, {G}AIA-C2-SP-OPM-FA-054,
  http://www.rssd.esa.int/doc\_fetch.php?id=2969346

\bibitem[{{Asiain}(1998)}]{1998PhDT........21A}
{Asiain}, R. 1998, PhD thesis, PhD Thesis, Universitat de Barcelona, Spain,
  (1998)

\bibitem[{{Asiain} {et~al.}(1997){Asiain}, {Torra}, \&
  {Figueras}}]{1997A&A...322..147A}
{Asiain}, R., {Torra}, J., \& {Figueras}, F. 1997, \aap, 322, 147

\bibitem[{{Balona} \& {Shobbrook}(1984)}]{1984MNRAS.211..375B}
{Balona}, L.~A. \& {Shobbrook}, R.~R. 1984, \mnras, 211, 375

\bibitem[{{Bertelli} {et~al.}(2008){Bertelli}, {Girardi}, {Marigo}, \&
  {Nasi}}]{2008A&A...484..815B}
{Bertelli}, G., {Girardi}, L., {Marigo}, P., \& {Nasi}, E. 2008, \aap, 484, 815

\bibitem[{{Bertelli} {et~al.}(2009){Bertelli}, {Nasi}, {Girardi}, \&
  {Marigo}}]{2009A&A...508..355B}
{Bertelli}, G., {Nasi}, E., {Girardi}, L., \& {Marigo}, P. 2009, \aap, 508, 355

\bibitem[{{Bessell} {et~al.}(1998){Bessell}, {Castelli}, \&
  {Plez}}]{1998A&A...333..231B}
{Bessell}, M.~S., {Castelli}, F., \& {Plez}, B. 1998, \aap, 333, 231

\bibitem[{{Bressan} {et~al.}(1993){Bressan}, {Fagotto}, {Bertelli}, \&
  {Chiosi}}]{1993A&AS..100..647B}
{Bressan}, A., {Fagotto}, F., {Bertelli}, G., \& {Chiosi}, C. 1993, \aaps, 100,
  647

\bibitem[{{Canuto} \& {Mazzitelli}(1991)}]{1991ApJ...370..295C}
{Canuto}, V.~M. \& {Mazzitelli}, I. 1991, \apj, 370, 295

\bibitem[{{Canuto} \& {Mazzitelli}(1992)}]{1992ApJ...389..724C}
{Canuto}, V.~M. \& {Mazzitelli}, I. 1992, \apj, 389, 724

\bibitem[{{Castelli} \& {Kurucz}(2004)}]{2004astro.ph..5087C}
{Castelli}, F. \& {Kurucz}, R.~L. 2004, ArXiv Astrophysics e-prints

\bibitem[{{Castelli} \& {Kurucz}(2006)}]{2006A&A...454..333C}
{Castelli}, F. \& {Kurucz}, R.~L. 2006, \aap, 454, 333

\bibitem[{{Claria Olmedo}(1974)}]{1974edfe.book.....C}
{Claria Olmedo}, J.~J. 1974, {Elementos de Fotometria ESTELAR}

\bibitem[{{Clem} {et~al.}(2004){Clem}, {VandenBerg}, {Grundahl}, \&
  {Bell}}]{2004AJ....127.1227C}
{Clem}, J.~L., {VandenBerg}, D.~A., {Grundahl}, F., \& {Bell}, R.~A. 2004, \aj,
  127, 1227

\bibitem[{{Crawford}(1978)}]{1978AJ.....83...48C}
{Crawford}, D.~L. 1978, \aj, 83, 48

\bibitem[{{Crawford}(1979)}]{1979AJ.....84.1858C}
{Crawford}, D.~L. 1979, \aj, 84, 1858

\bibitem[{{Crawford} \& {Mandwewala}(1976)}]{1976PASP...88..917C}
{Crawford}, D.~L. \& {Mandwewala}, N. 1976, \pasp, 88, 917

\bibitem[{{Dame} {et~al.}(2001){Dame}, {Hartmann}, \&
  {Thaddeus}}]{2001ApJ...547..792D}
{Dame}, T.~M., {Hartmann}, D., \& {Thaddeus}, P. 2001, \apj, 547, 792

\bibitem[{{Figueras} {et~al.}(1991){Figueras}, {Torra}, \&
  {Jordi}}]{1991A&AS...87..319F}
{Figueras}, F., {Torra}, J., \& {Jordi}, C. 1991, \aaps, 87, 319

\bibitem[{{Gonz{\'a}lez-Solares} {et~al.}(2008){Gonz{\'a}lez-Solares},
  {Walton}, {Greimel}, {Drew}, {Irwin}, {Sale}, {Andrews}, {Aungwerojwit},
  {Barlow}, {van den Besselaar}, {Corradi}, {G{\"a}nsicke}, {Groot}, {Hales},
  {Hopewell}, {Hu}, {Irwin}, {Knigge}, {Lagadec}, {Leisy}, {Lewis}, {Mampaso},
  {Matsuura}, {Moont}, {Morales-Rueda}, {Morris}, {Naylor}, {Parker}, {Prema},
  {Pyrzas}, {Rixon}, {Rodr{\'{\i}}guez-Gil}, {Roelofs}, {Sabin}, {Skillen},
  {Suso}, {Tata}, {Viironen}, {Vink}, {Witham}, {Wright}, {Zijlstra}, {Zurita},
  {Drake}, {Fabregat}, {Lennon}, {Lucas}, {Mart{\'{\i}}n}, {Phillipps},
  {Steeghs}, \& {Unruh}}]{2008MNRAS.388...89G}
{Gonz{\'a}lez-Solares}, E.~A., {Walton}, N.~A., {Greimel}, R., {et~al.} 2008,
  \mnras, 388, 89

\bibitem[{{Grosb{\o}l}(1978)}]{1978A&AS...32..409G}
{Grosb{\o}l}, P.~J. 1978, \aaps, 32, 409

\bibitem[{{Hauck} \& {Mermilliod}(1998)}]{1998A&AS..129..431H}
{Hauck}, B. \& {Mermilliod}, M. 1998, \aaps, 129, 431

\bibitem[{{Hilditch} {et~al.}(1983){Hilditch}, {Hill}, \&
  {Barnes}}]{1983MNRAS.204..241H}
{Hilditch}, R.~W., {Hill}, G., \& {Barnes}, J.~V. 1983, \mnras, 204, 241

\bibitem[{{Holmberg} {et~al.}(2007){Holmberg}, {Nordstr{\"o}m}, \&
  {Andersen}}]{2007A&A...475..519H}
{Holmberg}, J., {Nordstr{\"o}m}, B., \& {Andersen}, J. 2007, \aap, 475, 519

\bibitem[{{Karata{\c s}} \& {Schuster}(2010)}]{2010NewA...15..444K}
{Karata{\c s}}, Y. \& {Schuster}, W.~J. 2010, \na, 15, 444

\bibitem[{{Knude}(1978)}]{1978A&AS...33..347K}
{Knude}, J. 1978, \aaps, 33, 347

\bibitem[{{Kurucz}(1979)}]{1979ApJS...40....1K}
{Kurucz}, R. 1979, \apjs, 40, 1

\bibitem[{{Lindblad}(1967)}]{1967IAUS...31..143L}
{Lindblad}, P.~O. 1967, in IAU Symposium, Vol.~31, Radio Astronomy and the
  Galactic System, ed. H.~{van Woerden}, 143

\bibitem[{{Lindroos}(1980)}]{1980StoOR..17.....L}
{Lindroos}, K.~P. 1980, Stockholms Observatoriums Reports, 17

\bibitem[{{Lindroos}(1981)}]{1981StoOR..18.....L}
{Lindroos}, K.~P. 1981, Stockholms Obs. Rep., 18, 134

\bibitem[{{Luri} \& {Arenou}(1997)}]{1997ESASP.402..449L}
{Luri}, X. \& {Arenou}, F. 1997, in ESA Special Publication, Vol. 402,
  Hipparcos - Venice '97, ed. R.~M. {Bonnet}, E.~{H{\o}g}, P.~L. {Bernacca},
  L.~{Emiliani}, A.~{Blaauw}, C.~{Turon}, J.~{Kovalevsky}, L.~{Lindegren},
  H.~{Hassan}, M.~{Bouffard}, B.~{Strim}, D.~{Heger}, M.~A.~C. {Perryman}, \&
  L.~{Woltjer}, 449--452

\bibitem[{{Marco} {et~al.}(2001){Marco}, {Bernabeu}, \&
  {Negueruela}}]{2001AJ....121.2075M}
{Marco}, A., {Bernabeu}, G., \& {Negueruela}, I. 2001, \aj, 121, 2075

\bibitem[{{Mongui{\'o}} {et~al.}(2013){Mongui{\'o}}, {Figueras}, \&
  {Grosb{\o}l}}]{2013A&A...549A..78M}
{Mongui{\'o}}, M., {Figueras}, F., \& {Grosb{\o}l}, P. 2013, \aap, 549, A78

\bibitem[{{Moon} \& {Dworetsky}(1985)}]{1985MNRAS.217..305M}
{Moon}, T.~T. \& {Dworetsky}, M.~M. 1985, \mnras, 217, 305

\bibitem[{{Reis} \& {Corradi}(2008)}]{2008A&A...486..471R}
{Reis}, W. \& {Corradi}, W.~J.~B. 2008, \aap, 486, 471

\bibitem[{{Rieke} \& {Lebofsky}(1985)}]{1985ApJ...288..618R}
{Rieke}, G.~H. \& {Lebofsky}, M.~J. 1985, \apj, 288, 618

\bibitem[{{Roca-F{\`a}brega} {et~al.}(2013){Roca-F{\`a}brega}, {Valenzuela},
  {Figueras}, {Romero-G{\'o}mez}, {Vel{\'a}zquez}, {Antoja}, \&
  {Pichardo}}]{2013MNRAS.432.2878R}
{Roca-F{\`a}brega}, S., {Valenzuela}, O., {Figueras}, F., {et~al.} 2013,
  \mnras, 432, 2878

\bibitem[{{Russeil}(2003)}]{2003A&A...397..133R}
{Russeil}, D. 2003, \aap, 397, 133

\bibitem[{{Sale} {et~al.}(2010){Sale}, {Drew}, {Knigge}, {Zijlstra}, {Irwin},
  {Morris}, {Phillipps}, {Drake}, {Greimel}, {Unruh}, {Groot}, {Mampaso}, \&
  {Walton}}]{2010MNRAS.402..713S}
{Sale}, S.~E., {Drew}, J.~E., {Knigge}, C., {et~al.} 2010, \mnras, 402, 713

\bibitem[{{Schuster} \& {Nissen}(1989)}]{1989A&A...221...65S}
{Schuster}, W.~J. \& {Nissen}, P.~E. 1989, \aap, 221, 65

\bibitem[{{Sellwood}(2011)}]{2011MNRAS.410.1637S}
{Sellwood}, J.~A. 2011, \mnras, 410, 1637

\bibitem[{{Smalley} \& {Dworetsky}(1995)}]{1995A&A...293..446S}
{Smalley}, B. \& {Dworetsky}, M.~M. 1995, \aap, 293, 446

\bibitem[{{Smalley} \& {Kupka}(1997)}]{1997A&A...328..349S}
{Smalley}, B. \& {Kupka}, F. 1997, \aap, 328, 349

\bibitem[{{Str{\"o}mgren}(1966)}]{1966ARA&A...4..433S}
{Str{\"o}mgren}, B. 1966, \araa, 4, 433

\bibitem[{{Torra} {et~al.}(2000){Torra}, {Fern{\'a}ndez}, \&
  {Figueras}}]{2000A&A...359...82T}
{Torra}, J., {Fern{\'a}ndez}, D., \& {Figueras}, F. 2000, \aap, 359, 82

\bibitem[{{V{\'a}zquez} {et~al.}(2008){V{\'a}zquez}, {May}, {Carraro},
  {Bronfman}, {Moitinho}, \& {Baume}}]{2008ApJ...672..930V}
{V{\'a}zquez}, R.~A., {May}, J., {Carraro}, G., {et~al.} 2008, \apj, 672, 930

\end{thebibliography}

\Online
\begin{appendix}
\section{3D fitting algorithm}\label{Ap3Dfit}
In this section we describe the 3D fitting algorithm  developed to maximize the probability for one star to belong to a point of the theoretical grid in the $[m_1]-[c_1]-H\beta$ space.
We took into account the photometric errors in the three indexes that form the so-called ellipsoid of errors, as well as the 
distance between the star ($s$) and the point of the grid ($g$): 
\begin{equation}
D_{sg}=\sqrt{D_{sg,x}^2+D_{sg,y}^2+D_{sg,z}^2}=\xi-\xi_g \, ,
\end{equation}
with $D_{sg,x}=[c_1]_s-[c_1]_g$, $D_{sg,y}=[m_1]_s-[m_1]_g$, and $D_{sg,z}=[H\beta]_s-[H\beta]_g$ being the distances in each of the axes,
and $\xi-\xi_g$ being the distance along the axis between the star and the point of the grid.

\begin{figure}\centering
 \resizebox{8cm}{!}{\includegraphics{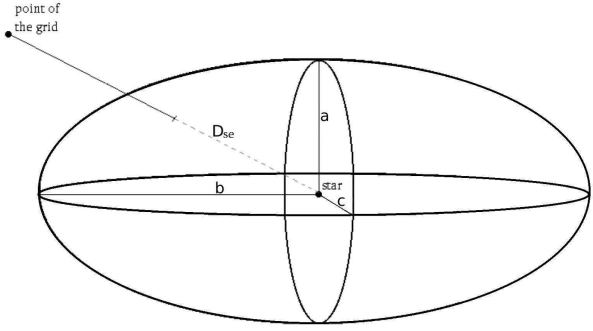}}
\caption{Scheme of the ellipsoid of errors that show how to compute the $\sigma_{sg}$ between the star and any point of the grid from the individual 
errors of the three photometric indexes.}
\label{ellip}
\end{figure}

The propagated photometric error in the direction between the star and the point of the grid 
($D_{se}$) is computed as the distance between the  location of the star and the surface of the ellipsoid of errors in the $\xi$ direction 
(see Fig. \ref{ellip}) 
with
$a=5\sigma_{[c_1]}$, $b=5\sigma_{[m_1]}$, and $c=5\sigma_{H\beta} $.
It is computed from the
equation of the ellipsoid and the equation of a line in the $\xi$ direction, that is,
\begin{eqnarray}\nonumber
\frac{1}{D_{se,x}^2}&=&\frac{1}{a^2}+\frac{1}{b^2}\cdot\frac{D_{sg,y}^2}{D_{sg,x}^2}+
\frac{1}{c^2}\cdot\frac{D_{sg,z}^2}{D_{sg,x}^2+D_{sg,y}^2}\cdot \left(1+\frac{D_{sg,y}^2}{D_{sg,x}^2}\right)\\ \nonumber
D_{se,y}^2&=&\frac{D_{sg,y}^2}{D_{sg,x}^2}\cdot D_{se,x}^2\\ \nonumber
D_{se,z}^2&=&D_{sg,z}^2\cdot\frac{D_{se,x}^2+D_{se,y}^2}{D_{sg,x}^2+D_{sg,y}^2}\,.
\end{eqnarray}
The 5$\sigma$ photometric error  between the star and the ellipsoid in the given direction is
\begin{equation}
D_{se}=\sqrt{D_{se,x}^2+D_{se,y}^2+D_{se,z}^2}\,.
\end{equation}

From this, the probability for one star to belong to a point of the grid is computed  centered on the star, with the standard deviation being $D_{se}$:
{\small
\begin{equation}
 P = 1 -  \frac{1}{D_{se}\sqrt{2\pi}}\left| \int_{-\infty}^{\xi_g} \exp \left[ \frac{(\xi-\xi_g)^2}{2D_{se}^2}\right] d\xi- \int_{\xi_g}^{\infty} \exp \left[ \frac{(\xi-\xi_g)^2}{2D_{se}^2}\right] d\xi  \right|\,.
\end{equation}}

\begin{figure}\centering
 \resizebox{\hsize}{!}{\includegraphics{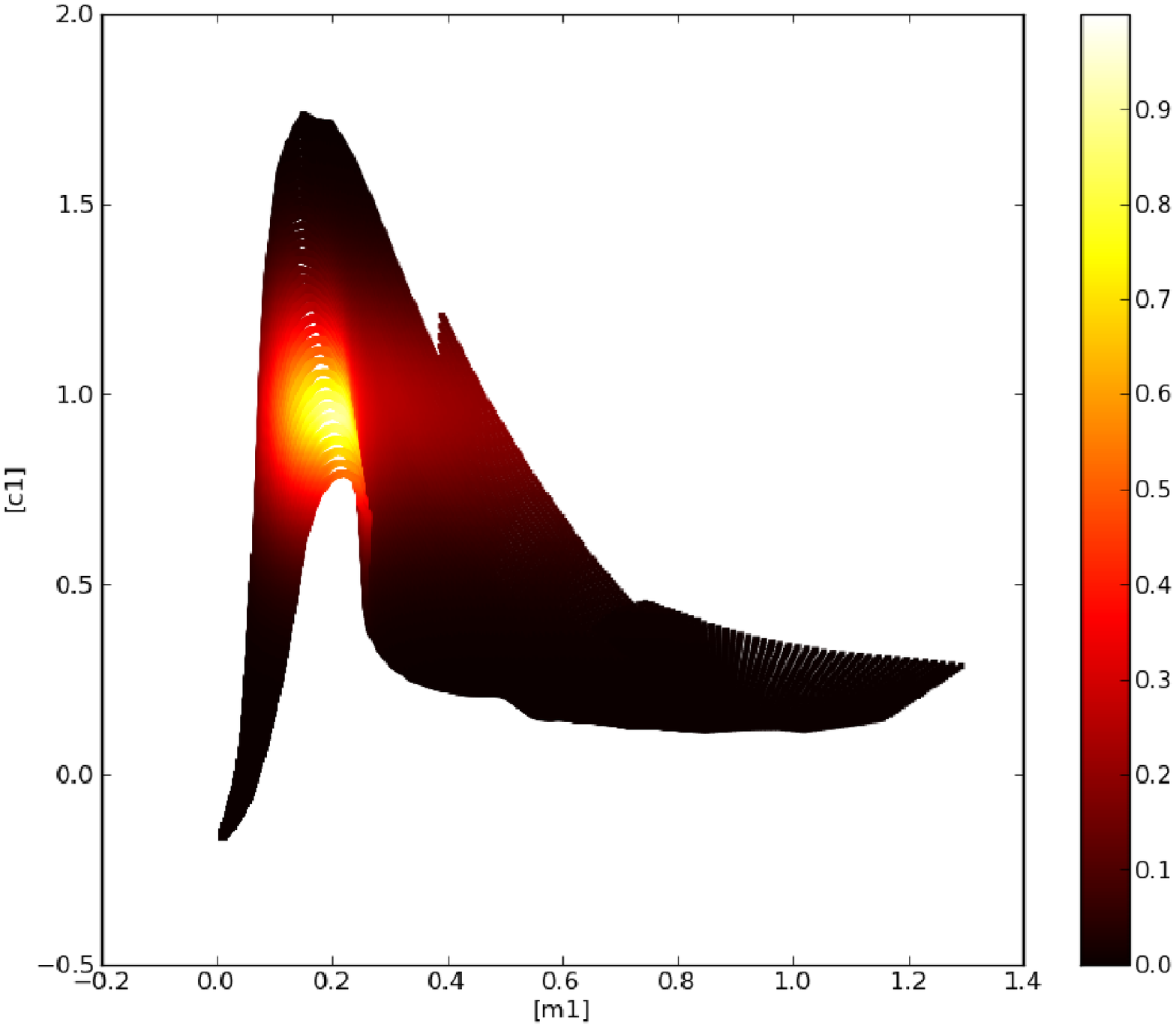}
  \includegraphics{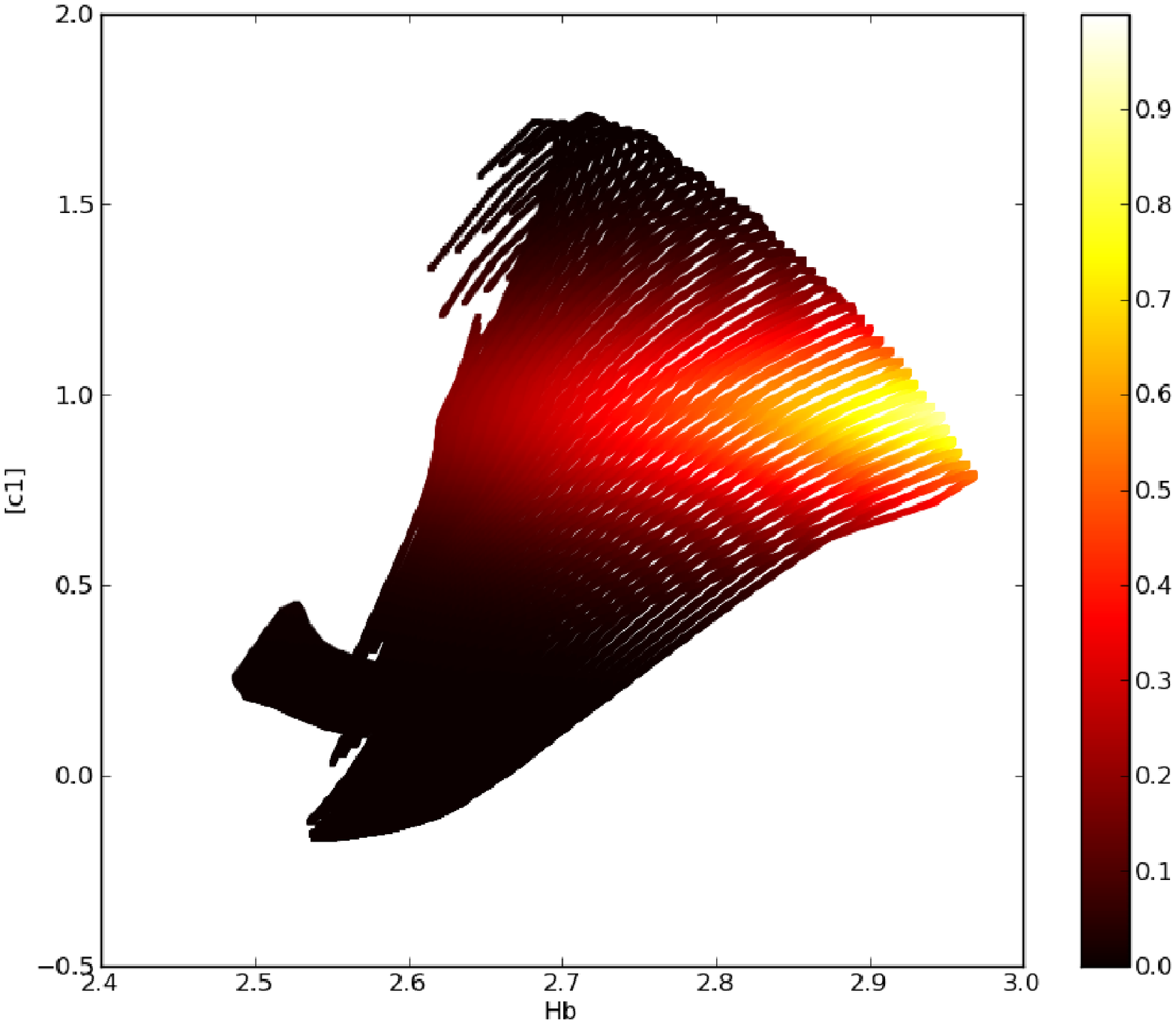}
}
\caption{$[c_1]$ - $[m_1]$ and a $[c_1]$ - $H\beta$ diagram. The color shows the corresponding probability $P$ for a star with ($[c_1]$, $[m_1]$, $H\beta$)=(0.96,0.15,2.96).}
\label{plotprob}
\end{figure}

For each star, we computed the probability $P$ for all the points of the grid to find the point with higher probability $P_{max}$. 
In Fig. \ref{plotprob} we show, for a single star, the probability for all the points of the grid in a $[c_1]$-$[m_1]$ and a $[c_1]$-$H\beta$ diagrams.
The original grids are discretized in steps of 0.5 or 0.25 in $\log g$ and 250\,K, 500\,K, or 1000\,K in $T_{eff}$.  
To develop our 3D fit the grids were interpolated in steps of 10\,K in $T_{eff}$ and 0.01 in $\log g$.

\end{appendix}
\begin{appendix}

\section{Binarity effect}\label{BinMet}

A significant fraction of the young stars in our survey can be binaries, either visual or physical. 
For physical binaries, some tests were developed to estimate the change in their photometric indexes and in turn the error introduced when ignoring binarity.
Their binarity ratio is debated, but according to \cite{LL:FA-054} 1) it can reach up to
80\% for the more massive stars,  
and 2) the mass ratio between the stars has a probability peak around $M_2/M_1$=0.6, with about 10\% of cases with a mass ratio higher than $M_2/M_1$=0.8.

\begin{figure}\centering
\resizebox{\hsize}{!}{\includegraphics{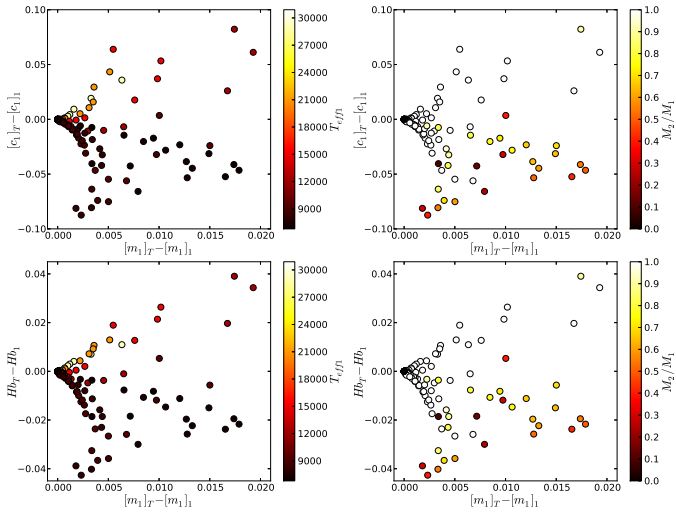}}
\caption{Differences in $[c_1]$ (top) and $H\beta$ (bottom) between the primary and the binary system. Left: the color shows the $T_{eff}$ of the primary. 
Right: the color shows the mass ratio between the primary and the secondary.}
\label{bin2}
\end{figure}

Simulations were made to estimate the change on the photometric indexes for different mass ratios. 
Different main-sequence-type stars from B0 to F0 were selected as primaries (checking all the cases, without considering the initial mass function). 
Then, for each primary, different secondaries were assumed, as well as their physical parameters.
We assigned to each star the corresponding photometric indexes ($(b-y)$, $m_1$, $c_1$, $H\beta$) according to the \cite{2004astro.ph..5087C} grids, and the assumed $T_{eff}$ and $\log g$=4.2.
Then the fluxes for the primary and the secondary stars were combined.
Figure \ref{bin2} shows the differences in the photometric indexes between the primary and the combined system 
for different $T_{eff}$ of the primary and different mass ratios. 
For primary stars with $T_{eff}>$7000\,K, the bias reaches values up to 0$\fm$08 in $[c_1]$,  up to 0$\fm$02 in $[m_1]$, and up to 0$\fm$04 in $H\beta$, 
which also leads to possible misclassifications.
As is known, the effect on absolute magnitude $M_V$ is highest when the two 
stars have equal luminosities, yielding an error of
0$\fm$75 (i.e., 30\% error in distance for this extreme case). 

\end{appendix}
\begin{appendix}

\section{New transformation coefficients}\label{trans2}
The transformation coefficients obtained for the new calibration after modifying the primary standard list are provided in Tables \ref{transcoef13} and \ref{transcoef13b}.
 These values have to be compared with those from Tables A.3 and A.4 from Paper I. Equation numbers refer to those from Paper I.
The new photometric indexes have also been uploaded to the CDS archive.

\begin{table*}\scriptsize
\caption{Standard transformation coefficients for the new calibration. Equation number refer to those from \cite{2013A&A...549A..78M}}\label{transcoef13}
\centering
\begin{tabular}
{c|cc|cc|ccc}%\hline
 & \multicolumn{2}{c|}{\textbf{Equation 2a}} &
\multicolumn{2}{c|}{\textbf{Equation 2b}} &
\multicolumn{3}{c}{\textbf{Equation 2c}} \\ \hline
 chip& $A_1$ & $B_1$ & $A_2$ & $C_2$ & $A_3$ & $B_3$ & $C_3$ \\ \hline
\multicolumn{8}{c}{\textbf{2009 Feb 13}} \\\hline
1&-24.915$\pm$0.002&-0.048$\pm$0.004&-0.227$\pm$0.003&0.977$\pm$0.006&-0.322$\pm$0.011&-0.082$\pm$0.016&0.952$\pm$0.009\\%\hline&
2&-24.716$\pm$0.002&-0.026$\pm$0.004&-0.274$\pm$0.002&0.987$\pm$0.004&-0.345$\pm$0.010&-0.092$\pm$0.013&0.976$\pm$0.009\\%\hline&
3&-24.857$\pm$0.002&-0.054$\pm$0.003&-0.238$\pm$0.002&0.992$\pm$0.003& 0.007$\pm$0.009&-0.104$\pm$0.011&0.956$\pm$0.009\\%\hline&
4&-24.802$\pm$0.001&-0.054$\pm$0.003&-0.221$\pm$0.002&0.978$\pm$0.003&-0.194$\pm$0.008&-0.057$\pm$0.012&0.988$\pm$0.007\\\hline
\multicolumn{8}{c}{\textbf{2009 Feb 16}}\\\hline
1&-24.834$\pm$0.001&-0.044$\pm$0.002&-0.247$\pm$0.002&0.975$\pm$0.003&-0.247$\pm$0.005&-0.139$\pm$0.006&0.974$\pm$0.005\\%\hline&
2&-24.634$\pm$0.002&-0.025$\pm$0.003&-0.297$\pm$0.001&1.000$\pm$0.002&-0.245$\pm$0.006&-0.180$\pm$0.007&0.962$\pm$0.006\\%\hline&
3&-24.770$\pm$0.001&-0.063$\pm$0.002&-0.257$\pm$0.002&0.984$\pm$0.003& 0.083$\pm$0.006&-0.114$\pm$0.007&0.979$\pm$0.005\\%\hline&
4&-24.717$\pm$0.002&-0.063$\pm$0.003&-0.253$\pm$0.002&0.995$\pm$0.003&-0.096$\pm$0.007&-0.100$\pm$0.010&0.974$\pm$0.006\\\hline
\multicolumn{8}{c}{\textbf{2011 Jan 08}}\\\hline
1&-24.815$\pm$0.001&-0.045$\pm$0.002&-0.218$\pm$0.001&0.967$\pm$0.002&-0.313$\pm$0.005&-0.133$\pm$0.005&0.986$\pm$0.005\\%\hline
2&-24.601$\pm$0.001&-0.025$\pm$0.002&-0.269$\pm$0.002&1.006$\pm$0.003&-0.316$\pm$0.004&-0.135$\pm$0.005&0.960$\pm$0.004\\%\hline
3&-24.699$\pm$0.001&-0.061$\pm$0.002&-0.219$\pm$0.001&0.984$\pm$0.002&-0.030$\pm$0.005&-0.120$\pm$0.006&0.975$\pm$0.005\\%\hline
4&-24.693$\pm$0.001&-0.065$\pm$0.002&-0.228$\pm$0.002&0.999$\pm$0.003&-0.176$\pm$0.007&-0.087$\pm$0.008&0.954$\pm$0.006\\\hline
\multicolumn{8}{c}{\textbf{2011 Jan 09}}\\\hline
1&-24.741$\pm$0.004&-0.047$\pm$0.011&-0.235$\pm$0.005&0.970$\pm$0.014&-0.330$\pm$0.011&-0.127$\pm$0.027&0.992$\pm$0.012\\%\hline
2&-24.523$\pm$0.004&-0.027$\pm$0.011&-0.286$\pm$0.005&1.004$\pm$0.013&-0.339$\pm$0.014&-0.148$\pm$0.031&0.966$\pm$0.014\\%\hline
3&-24.625$\pm$0.008&-0.060$\pm$0.022&-0.238$\pm$0.007&0.981$\pm$0.019&-0.053$\pm$0.011&-0.105$\pm$0.027&0.981$\pm$0.012\\%\hline
4&-24.618$\pm$0.005&-0.064$\pm$0.011&-0.248$\pm$0.005&1.001$\pm$0.013&-0.208$\pm$0.014&-0.085$\pm$0.029&0.961$\pm$0.013\\\hline
\multicolumn{8}{c}{\textbf{2011 Jan 10}}\\\hline
1&-24.764$\pm$0.001&-0.048$\pm$0.002&-0.225$\pm$0.001&0.971$\pm$0.003&-0.256$\pm$0.004&-0.114$\pm$0.005&0.972$\pm$0.004\\%\hline
2&-24.550$\pm$0.001&-0.031$\pm$0.002&-0.277$\pm$0.001&0.996$\pm$0.002&-0.251$\pm$0.005&-0.127$\pm$0.006&0.957$\pm$0.004\\%\hline
3&-24.650$\pm$0.001&-0.055$\pm$0.002&-0.235$\pm$0.001&0.988$\pm$0.003& 0.020$\pm$0.005&-0.110$\pm$0.006&0.968$\pm$0.005\\%\hline
4&-24.646$\pm$0.001&-0.058$\pm$0.002&-0.229$\pm$0.001&0.988$\pm$0.003&-0.136$\pm$0.005&-0.089$\pm$0.007&0.961$\pm$0.005\\\hline
\multicolumn{8}{c}{\textbf{2011 Jan 11}}\\\hline
1&-24.720$\pm$0.002&-0.037$\pm$0.003&-0.232$\pm$0.002&0.970$\pm$0.003&-0.279$\pm$0.005&-0.123$\pm$0.006&0.981$\pm$0.006\\%\hline
2&-24.501$\pm$0.002&-0.019$\pm$0.004&-0.285$\pm$0.002&1.006$\pm$0.003&-0.358$\pm$0.006&-0.141$\pm$0.007&0.964$\pm$0.004\\%\hline
3&-24.600$\pm$0.002&-0.052$\pm$0.003&-0.235$\pm$0.002&0.983$\pm$0.003& 0.005$\pm$0.006&-0.119$\pm$0.007&0.972$\pm$0.006\\%\hline
4&-24.600$\pm$0.002&-0.049$\pm$0.004&-0.233$\pm$0.002&0.993$\pm$0.003&-0.157$\pm$0.006&-0.082$\pm$0.008&0.956$\pm$0.005\\\hline
\multicolumn{8}{c}{\textbf{2011 Feb 16}}\\\hline
1&-24.748$\pm$0.003&-0.051$\pm$0.006&-0.241$\pm$0.005&0.988$\pm$0.009&-0.342$\pm$0.013&-0.085$\pm$0.019&0.995$\pm$0.012\\%\hline
2&-24.528$\pm$0.003&-0.026$\pm$0.005&-0.291$\pm$0.004&0.994$\pm$0.007&-0.308$\pm$0.013&-0.145$\pm$0.017&0.941$\pm$0.009\\%\hline
3&-24.623$\pm$0.003&-0.065$\pm$0.005&-0.245$\pm$0.004&0.986$\pm$0.007&-0.046$\pm$0.013&-0.092$\pm$0.020&0.973$\pm$0.011\\%\hline
4&-24.638$\pm$0.003&-0.037$\pm$0.006&-0.235$\pm$0.005&0.982$\pm$0.011&-0.205$\pm$0.017&-0.094$\pm$0.028&0.977$\pm$0.012\\\hline
\multicolumn{8}{c}{\textbf{2011 Feb 17}}\\\hline
1&-24.769$\pm$0.003&-0.061$\pm$0.005&-0.217$\pm$0.003&0.962$\pm$0.006&-0.318$\pm$0.008&-0.127$\pm$0.011&0.989$\pm$0.009\\%\hline
2&-24.554$\pm$0.003&-0.029$\pm$0.005&-0.293$\pm$0.004&0.989$\pm$0.007&-0.331$\pm$0.010&-0.134$\pm$0.012&0.947$\pm$0.007\\%\hline
3&-24.662$\pm$0.003&-0.059$\pm$0.004&-0.230$\pm$0.003&0.969$\pm$0.006&-0.033$\pm$0.011&-0.125$\pm$0.014&0.960$\pm$0.009\\%\hline
4&-24.665$\pm$0.003&-0.050$\pm$0.005&-0.233$\pm$0.003&0.998$\pm$0.006&-0.217$\pm$0.010&-0.074$\pm$0.015&0.987$\pm$0.008\\%\hline
\end{tabular}
\end{table*}

\begin{table*}\scriptsize
\caption{Standard transformation coefficients for the new calibration (Cont.)}\label{transcoef13b}
\centering
\begin{tabular}
{c|cccc|cc|ccc}%\hline
 & \multicolumn{4}{c|}{\textbf{Equation 2d}} & 
\multicolumn{2}{c|}{\textbf{Equation 3b}}&
\multicolumn{3}{c}{\textbf{Equation  3a}}\\ \hline
 chip& $A_4$ & $B_4$ & $C_4$ & $D_4$ & $\tilde{A_5}$ & $\tilde{C_5}$ & $\tilde{A_4}$ & $\tilde{B_4}$ & $\tilde{C_4}$ \\ \hline
\multicolumn{10}{c}{\textbf{2009 Feb 13}} \\\hline
1& 0.173$\pm$0.006&0.229$\pm$0.016&0.878$\pm$0.012&0.033$\pm$0.005&2.324$\pm$0.001&0.962$\pm$0.004&0.203$\pm$0.004&0.241$\pm$0.016&0.849$\pm$0.012\\%\hline&
2& 0.230$\pm$0.005&0.283$\pm$0.014&0.886$\pm$0.011&0.028$\pm$0.005&2.308$\pm$0.001&0.910$\pm$0.004&0.254$\pm$0.004&0.294$\pm$0.015&0.864$\pm$0.010\\%\hline&
3& 0.294$\pm$0.005&0.235$\pm$0.012&0.891$\pm$0.008&0.023$\pm$0.004&2.331$\pm$0.001&0.998$\pm$0.003&0.314$\pm$0.003&0.224$\pm$0.012&0.887$\pm$0.008\\%\hline&
4& 0.352$\pm$0.004&0.326$\pm$0.011&0.834$\pm$0.008&0.023$\pm$0.003&2.322$\pm$0.001&0.970$\pm$0.003&0.371$\pm$0.003&0.317$\pm$0.011&0.830$\pm$0.008\\\hline
\multicolumn{10}{c}{\textbf{2009 Feb 16}}\\\hline
1& 0.133$\pm$0.005&0.263$\pm$0.010&0.868$\pm$0.009&0.009$\pm$0.005&2.311$\pm$0.001&0.976$\pm$0.005&0.140$\pm$0.003&0.270$\pm$0.010&0.859$\pm$0.008\\%\hline&
2& 0.168$\pm$0.003&0.308$\pm$0.004&0.891$\pm$0.004&0.033$\pm$0.003&2.283$\pm$0.001&0.931$\pm$0.005&0.196$\pm$0.002&0.332$\pm$0.004&0.858$\pm$0.003\\%\hline&
3& 0.254$\pm$0.004&0.235$\pm$0.008&0.894$\pm$0.007&0.014$\pm$0.003&2.322$\pm$0.001&1.023$\pm$0.005&0.266$\pm$0.002&0.241$\pm$0.008&0.883$\pm$0.006\\%\hline&
4& 0.305$\pm$0.004&0.316$\pm$0.009&0.857$\pm$0.007&0.023$\pm$0.003&2.304$\pm$0.001&0.961$\pm$0.005&0.327$\pm$0.002&0.320$\pm$0.010&0.841$\pm$0.006\\\hline
\multicolumn{10}{c}{\textbf{2011 Jan 08}}\\\hline
1& 0.187$\pm$0.003&0.234$\pm$0.006&0.891$\pm$0.005&0.011$\pm$0.003&2.318$\pm$0.001&0.989$\pm$0.006&0.196$\pm$0.002&0.880$\pm$0.004&0.249$\pm$0.005\\%\hline
2& 0.233$\pm$0.003&0.273$\pm$0.004&0.894$\pm$0.003&0.031$\pm$0.002&2.300$\pm$0.001&0.946$\pm$0.006&0.259$\pm$0.002&0.865$\pm$0.003&0.295$\pm$0.004\\%\hline
3& 0.301$\pm$0.003&0.212$\pm$0.005&0.914$\pm$0.005&0.017$\pm$0.003&2.304$\pm$0.001&1.043$\pm$0.005&0.316$\pm$0.002&0.899$\pm$0.004&0.220$\pm$0.005\\%\hline
4& 0.351$\pm$0.003&0.311$\pm$0.006&0.857$\pm$0.004&0.029$\pm$0.002&2.302$\pm$0.001&0.963$\pm$0.005&0.378$\pm$0.002&0.838$\pm$0.005&0.314$\pm$0.007\\\hline
\multicolumn{10}{c}{\textbf{2011 Jan 09}}\\\hline
1& 0.168$\pm$0.005&0.221$\pm$0.027&0.912$\pm$0.019&0.015$\pm$0.006&2.316$\pm$0.002&0.978$\pm$0.018&0.174$\pm$0.005&0.214$\pm$0.027&0.919$\pm$0.019\\%\hline
2& 0.216$\pm$0.006&0.267$\pm$0.026&0.905$\pm$0.016&0.036$\pm$0.005&2.306$\pm$0.003&0.955$\pm$0.020&0.238$\pm$0.005&0.232$\pm$0.028&0.922$\pm$0.017\\%\hline
3& 0.280$\pm$0.005&0.205$\pm$0.026&0.926$\pm$0.017&0.025$\pm$0.006&2.314$\pm$0.002&1.038$\pm$0.018&0.290$\pm$0.005&0.205$\pm$0.027&0.930$\pm$0.017\\%\hline
4& 0.344$\pm$0.005&0.287$\pm$0.025&0.871$\pm$0.017&0.029$\pm$0.006&2.302$\pm$0.003&0.969$\pm$0.020&0.356$\pm$0.005&0.290$\pm$0.027&0.872$\pm$0.018\\\hline
\multicolumn{10}{c}{\textbf{2011 Jan 10}}\\\hline
1& 0.165$\pm$0.003&0.219$\pm$0.006&0.903$\pm$0.005&0.021$\pm$0.003&2.322$\pm$0.001&0.966$\pm$0.005&0.182$\pm$0.002&0.226$\pm$0.006&0.889$\pm$0.005\\%\hline
2& 0.203$\pm$0.002&0.270$\pm$0.006&0.892$\pm$0.005&0.036$\pm$0.002&2.313$\pm$0.001&0.938$\pm$0.006&0.234$\pm$0.002&0.294$\pm$0.007&0.857$\pm$0.005\\%\hline
3& 0.269$\pm$0.003&0.227$\pm$0.006&0.905$\pm$0.005&0.028$\pm$0.003&2.320$\pm$0.001&1.027$\pm$0.005&0.293$\pm$0.002&0.231$\pm$0.007&0.888$\pm$0.005\\%\hline
4& 0.338$\pm$0.003&0.306$\pm$0.007&0.843$\pm$0.005&0.024$\pm$0.003&2.306$\pm$0.001&0.957$\pm$0.005&0.359$\pm$0.002&0.308$\pm$0.007&0.829$\pm$0.005\\\hline
\multicolumn{10}{c}{\textbf{2011 Jan 11}}\\\hline
1& 0.162$\pm$0.003&0.231$\pm$0.006&0.894$\pm$0.006&0.013$\pm$0.003&2.329$\pm$0.002&0.946$\pm$0.010&0.172$\pm$0.002&0.239$\pm$0.006&0.883$\pm$0.005\\%\hline
2& 0.204$\pm$0.003&0.250$\pm$0.007&0.914$\pm$0.006&0.033$\pm$0.003&2.315$\pm$0.001&0.941$\pm$0.007&0.232$\pm$0.002&0.291$\pm$0.008&0.867$\pm$0.006\\%\hline
3& 0.273$\pm$0.003&0.214$\pm$0.006&0.917$\pm$0.005&0.020$\pm$0.003&2.322$\pm$0.001&1.035$\pm$0.007&0.290$\pm$0.002&0.227$\pm$0.006&0.899$\pm$0.004\\%\hline
4& 0.333$\pm$0.003&0.295$\pm$0.007&0.859$\pm$0.006&0.026$\pm$0.003&2.305$\pm$0.001&0.963$\pm$0.007&0.357$\pm$0.002&0.310$\pm$0.008&0.833$\pm$0.006\\\hline
\multicolumn{10}{c}{\textbf{2011 Feb 16}}\\\hline
1& 0.190$\pm$0.008&0.187$\pm$0.016&0.919$\pm$0.013&0.013$\pm$0.007&2.302$\pm$0.002&0.922$\pm$0.014&0.201$\pm$0.005&0.189$\pm$0.016&0.910$\pm$0.012\\%\hline
2& 0.226$\pm$0.007&0.293$\pm$0.017&0.879$\pm$0.013&0.039$\pm$0.005&2.289$\pm$0.001&0.918$\pm$0.010&0.262$\pm$0.005&0.336$\pm$0.018&0.827$\pm$0.012\\%\hline
3& 0.291$\pm$0.007&0.204$\pm$0.017&0.919$\pm$0.014&0.023$\pm$0.006&2.302$\pm$0.002&1.024$\pm$0.012&0.312$\pm$0.005&0.217$\pm$0.017&0.899$\pm$0.013\\%\hline
4& 0.341$\pm$0.009&0.271$\pm$0.021&0.893$\pm$0.017&0.033$\pm$0.006&2.279$\pm$0.002&0.955$\pm$0.011&0.376$\pm$0.006&0.303$\pm$0.021&0.842$\pm$0.015\\\hline
\multicolumn{10}{c}{\textbf{2011 Feb 17}}\\\hline
1& 0.209$\pm$0.005&0.210$\pm$0.010&0.915$\pm$0.009&0.019$\pm$0.006&2.301$\pm$0.002&0.959$\pm$0.011&0.223$\pm$0.003&0.212$\pm$0.010&0.906$\pm$0.008\\%\hline
2& 0.270$\pm$0.006&0.356$\pm$0.016&0.832$\pm$0.012&0.033$\pm$0.006&2.292$\pm$0.001&0.924$\pm$0.007&0.298$\pm$0.005&0.395$\pm$0.015&0.788$\pm$0.010\\%\hline
3& 0.319$\pm$0.006&0.212$\pm$0.012&0.934$\pm$0.009&0.030$\pm$0.005&2.311$\pm$0.001&1.030$\pm$0.008&0.343$\pm$0.004&0.236$\pm$0.013&0.905$\pm$0.009\\%\hline
4& 0.399$\pm$0.006&0.261$\pm$0.014&0.878$\pm$0.012&0.023$\pm$0.005&2.289$\pm$0.001&0.976$\pm$0.008&0.421$\pm$0.004&0.279$\pm$0.014&0.850$\pm$0.011\\%\hline
\end{tabular}
\end{table*}

\end{appendix}
\begin{appendix}

\section{Catalog of young stars}\label{CAT}
In Table \ref{header} we describe the columns available for the SPP catalog of our anticenter survey. The complete catalog can be found at the CDS data archive. First an identifier and RA-DEC coordinates for the star are given, 
as well as the photometric indexes (with the new value of $H\beta$) and their errors. 
Then all the SPP from the MB and EC methods are listed, also with the errors and flags.
For the MB, the (B) SPP obtained from the second maxim probability at the other side of the gap between early- 
and late-type regions are also provided.
Only the 13687 stars with either $T_{eff}>$7000\,K from MB or classified as O-A9 by EC methods are included in the catalog.
The physical parameters for stars with  $T_{eff}<$7000\,K (either as a secondary (B) SPP or for stars differently classified by the two methods) are only tentative, 
and they must be used with caution.
\begin{table*}\scriptsize
\caption{Description of the columns available for the SPP catalog}\label{header}
\centering
\begin{tabular} {l|lcl}\hline
\textbf{Column} & \textbf{Label} & \textbf{Units} & \textbf{Description} \\\hline
1 & ID       & -&ID number\\ %\hline
2 & RAdeg     & deg  & Right ascension J2000.0 \\ %\hline
3 & DEdeg      & deg & Declination J2000.0 \\ %\hline
4 & Vmag       & mag & Magnitude transformed into the standard Johnson V magnitude \\% \hline
5 & e\_Vmag   & mag & Error of Vmag\\ %\hline
6 & (b-y)    & mag   & Str\"{o}mgren (b-y) color index \\ %\hline
7 & e\_(b-y) & mag &Error of (b-y)\\%\hline
8 & c1  & mag     & Str\"{o}mgren $c_1$ index \\ %\hline
9 & e\_c1 &mag & Error of c1\\%\hline
10& m1 & mag     & Str\"{o}mgren $m_1$ index \\%\hline
11& e\_m1 & mag &Error of m1\\%\hline
12& Hbeta & mag      & New values for the Str\"{o}mgren H$\beta$ index \\% \hline
13& e\_Hbeta & mag &Error of Hbeta\\\hline

14&r\_MB&pc&MB distance\\%\hline
15&e\_r\_MB&pc&Error of r\_MB\\%\hline
16&(b-y)0\_MB&mag&MB intrinsic color $(b-y)_0$\\%\hline
17&e\_(b-y)0\_MB&mag&Error of (b-y)0\_MB\\%\hline
18&Mv\_MB&mag&MB absolute magnitude\\%\hline
19&e\_Mv\_MB&mag&Error of Mv\_MB\\%\hline
20&Av\_MB&mag&MB visual absorption\\%\hline
21&e\_Av\_MB&mag&Error of Av\_MB\\%\hline
22&Teff&K&MB effective temperature\\%\hline
23&e\_Teff&K&Error of Teff\\%\hline
24&logg&-&MB surface gravity\\%\hline
25&e\_logg&-&Error of logg\\%\hline
26&lum&$L_{\odot}$&MB luminosity\\%\hline
27&e\_lum&$L_{\odot}$&Error of lum\\%\hline
28&logAge&-&MB $\log$(Age) \\%\hline
29&e\_logAge&-&Error of logAge\\%\hline
30&Mass&$M_{\odot}$&MB stellar mass\\%\hline
31&e\_Mass&$M_{\odot}$&Error of Mass\\%\hline
32&BC&mag&MB bolometric correction\\%\hline
33&e\_BC&mag&Error of BC\\%\hline
34&DSE&-&$D_{se}$: 3D fitting parameter providing the distance between the star and the ellipsoid of errors in the $[m_1]-[c_1]-H\beta$ space\\%\hline
35&DSG&-&$D_{sg}$: 3D fitting parameter providing the distance between the star and the point if the grid in the $[m_1]-[c_1]-H\beta$ space\\%\hline
36&Pmax&-&$P_{max}$ parameter indicating the probability for the star to belong to the most probable point of the grid\\%\hline
37&PmaxB&-&$P_{max}^B$ parameter indicating the probability for the star to belong to the corresponding point of the grid at the other side of the gap\\%\hline
38&Nside&-&$N_{side}$: \% of simulated stars located at the corresponding side of the gap $^{(1)}$\\\hline

39&r\_EC&pc&EC distance\\%\hline
40&e\_r\_EC&pc&Error of r\_EC\\%\hline
41&(b-y)0\_EC&mag&EC intrinsic color $(b-y)_0$\\%\hline
42&e\_(b-y)0\_EC&mag&Error of (b-y)0\_EC\\%\hline
43&Mv\_EC&mag&EC absolute magnitude\\%\hline
44&e\_Mv\_EC&mag&Error of Mv\_EC\\%\hline
45&Av\_EC&mag&EC visual absorption \\%\hline
46&e\_Av\_EC&mag&Error of Av\_EC\\%\hline
47&regNC&-&Photometric region assigned to the star by the NC method\\% \hline
48&Nreg1&-& \% of simulated stars located in the early region (up to B9)\\%\hline
49&Nreg2&-&\% of simulated stars located in the intermediate region (A0-A3)\\%\hline
50&Nreg3&-&\% of simulated stars located in the late region (up to A9)\\%\hline
51&Nreg4&-&\% of simulated stars located in the late region (F0-G2)\\%\hline
52&Nreg5&-&\% of simulated stars located in the late region (later than G2)\\\hline

53&rB&pc&MB (B) distance\\%\hline
54&(b-y)0B&mag&MB (B) $(b-y)_0$ intrinsic color \\%\hline
55&MvB&mag&MB (B) absolute magnitude\\%\hline
56&AvB&mag&MB (B) visual absorption\\%\hline
57&TeffB&K&MB (B) effective temperature\\%\hline
58&loggB&-&MB (B) surface gravity\\%\hline
59&lumB&$L_{\odot}$&MB (B) luminosity\\%\hline
60&logAgeB&-&MB (B) $\log$(Age)\\%\hline
61&MassB&$M_{\odot}$&MB (B) stellar mass\\%\hline
62&BCB&mag& MB (B) bolometric correction\\\hline

63&EMLS&-&Flag indicating emission line stars according to IPHAS data $^{(2)}$  \\%\hline
64&DJKmin&mag&Minimum distance between the location of the star in the $(J-K)_0$ vs. $(b-y)_0$ plot and the main sequence relation (MB method)\\%\hline
65&DKHin&mag&Minimum distance between the location of the star in the $(J-H)_0$ vs. $(b-y)_0$ plot and the main sequence relation (MB method)\\%\hline
66&FlagJK&-&$Flag_{JK}=D_{JK,B,min}-D_{JK,min}$ comparing both assignations with 2MASS data\\%\hline
67&FlagJH&-&$Flag_{JH}=D_{JH,B,min}-D_{JH,min}$ comparing both assignations with 2MASS data
\end{tabular}
\tablefoot{
(B) Indicates second option parameters forced to be at the other side of the $[c_1]-[m_1]$ gap.\\
(1) The limit between sides is empirically set at 10000\,K. \\
(2) 0: if it is not an emission line star, 1: if it is an emission line star, 2: when IPHAS information is not available.}
\end{table*}
\end{appendix}

\end{document}